\documentclass[]{aa}

\usepackage[varg]{txfonts}
\usepackage{graphicx,rotating,natbib,color}
\usepackage{amsmath,amssymb}

\begin{document}

\title{Thermal history modeling of the L chondrite parent body}


\author{
 Hans-Peter Gail\inst{1}
 \and Mario Trieloff\inst{2}
}

\institute{
Zentrum f\"ur Astronomie, Institut f\"ur Theoretische Astrophysik, 
           Universit\"at Heidelberg, Albert-Ueberle-Str. 2,
           69120 Heidelberg, Germany, \email{gail@uni-heidelberg.de}
\and
Klaus-Tschira-Labor f\"ur Kosmochemie, Institut f\"ur Geowissenschaften, Universit\"at Heidelberg, Im Neuenheimer Feld 236, 69120 Heidelberg, Germany, \email{Mario.Trieloff@geow.uni-heidelberg.de}
  }

\date{Received date ; accepted date}

\abstract
{
The cooling history of individual meteorites can be reconstructed if closure temperatures and closure ages of different radioisotopic chronometers are available for a couple of meteorites. If a high similarity of chemical and isotopic composition suggests a common origin from the same parent body, some basic properties of this body can be derived. 
}
{
The radius of the L chondrite parent body, its formation time, and its evolution history are determined by fitting theoretical models to empirical data of radioisotopic chronometers for L chondrites. 
}
{
A simplified evolution model for the L chondrite parent body is constructed considering sintering of the initially porous material, temperature dependent heat conductivity, and an insulating regolith layer. Such models are fitted to thermochronological data of five meteorites for which precise data for the Hf-W and U-Pb-Pb thermochronometers have been published.
}
{
A set of parameters for the L chondrite parent body is found that yields excellent agreement (within error bounds) between a thermal evolution model and thermochonological data of five examined L chondrites. Empirical cooling rate data also agree with the model results within error bounds such that there is no conflict between cooling rate data and the onion-shell model. Two models are found to be compatible with the presently available empirical data: One model with a radius of 115 km and a formation time of 1.89 Ma after CAI formation, another model with 160 km radius and formation time of 1.835 Ma. The central temperature of the smaller body remains well below the Ni,Fe-FeS eutectic melting temperature and is consistent with the apparent non-existence of primitive achondrites related to the L chondrites. For the bigger model incipient melting in the central core region is predicted which opens the possibility that primitive achondrites related to L chondrites could be found.
}
{}


\maketitle



\section{Introduction}

Radiometric ages for chondritic meteorites and their components provide information on the accretion timescale of chondrite parent bodies, and on the cooling history within certain regions of these bodies. However, to apply this age information for constraining the internal structure, and the accretion and cooling history of the chondrite parent bodies, the empirical cooling paths obtained by dating chondrites must be combined with theoretical models of the thermal evolution of planetesimals. Important parameters in such thermal models include the initial abundances of heat-producing short-lived radio nuclides ($^{26\!}$Al and $^{60}$Fe), which depend on accretion time of the parent body, the terminal size of the parent body, and the chemical composition and physical properties of the chondritic material. Part of these parameters, like material properties, can be determined from laboratory investigations of meteorites, and others, like accretion time and radius, have to be found by comparing evolution models with empirical cooling histories of the meteorites. For L chondrites and their parent body this has been done in various degrees of approximation by \citep{Miy81, Ben95,Ben96,Ben02,Bou07,Spr11, Mar14,Bla17}. Here we use the more detailed method for modelling the structure and thermal evolution of planetesimals of $\approx100$\,km size of \citet[][henceforth called paper I]{Hen11} and its extensions \citep{Hen12,Hen13, Gai15,Gai18} to reconstruct the properties of the L chondrite parent body, and to evaluate the most important parameters that determined its thermal evolution. 

\begin{table*}[t]
\caption{Cooling ages (in Ma) determined for L chondrites. Data of meteorites marked with a hook are used for model optimisation.}

{\small
\begin{tabular*}{\hsize}{lll@{\,}lr@{\,$\pm$\,}l@{\,}lr@{$\,\pm\,$}l@{\,}lr@{\,$\pm$\,}l@{\,}lr@{\,$\pm$\,}l@{\,}lc}
\hline
\noalign{\smallskip}
&  & \multicolumn{2}{c}{Shock-} & \multicolumn{11}{c}{Thermochronometer} \\
& Type & \multicolumn{2}{c}{stage} & \multicolumn{3}{c}{Hf-W} & \multicolumn{3}{c}{U-Pb-Pb}  & \multicolumn{3}{c}{I-Xe} & \multicolumn{3}{c}{Ar-Ar$^{\rm d}$} & Used \\
\noalign{\smallskip}
\hline
\noalign{\smallskip}
Closure-&  &  &  & 1148 & 50  & &  720 & 50 & &  770$^{\rm a}$ & 60 & & 550 & 20 &  \\
temperature [K] &  &  &  &  \multicolumn{3}{c}{\quad} & \multicolumn{3}{c}{\quad} & 650$^{\rm b}$ & 50 & \\
\noalign{\bigskip}
Bruderheim     & L6     & S2/S4 & (15,16) & 4556.9 & 0.8 & (1) & 4514.58 & 0.54 & (1) & 4500.$^{\rm c}$\  & & (9) & \multicolumn{3}{c}{\quad}  & $\surd$ \\
               &      &    &    & \multicolumn{3}{c}{\quad} &  4509.55 & 1.36 & (6) \\
Saratov        & L4     & S2/S3 & (15,17,18) & 4563.9 & 0.7 & (8) & 4530.0  & 6.0 & (4) & 4551.0$^{\rm c}$ & 0.9 & (12) & 4440.0 & 30.0 & (5) & \\
Elenovka       & L5     & S2 & (16) & 4558.8 & 0.8 & (8) & 4535.0  & 1.0 & (3) &   \multicolumn{3}{c}{\quad}  &  \multicolumn{3}{c}{\quad} & $\surd$ \\
               &  \multicolumn{3}{c}{\quad} &   \multicolumn{3}{c}{\quad}  & 4555.03 & 0.52 & (11) & 4550$^{\rm c}$ & 5  & (11) \\
Homestead      & L5     &   &   & 4562.8 & 1.3  & (8) & 4514.5  & 1.3 & (2) &   \multicolumn{3}{c}{\quad}  &   \multicolumn{3}{c}{\quad}  & \\
Barwell        & L5-6   &   &   & 4559.7 & 0.7 & (8) & 4538.4  & 0.8 & (2) & \multicolumn{3}{c}{\quad} & 4430.0 & 40.0 & (5) & $\surd$ \\
               &        &   & & \multicolumn{3}{c}{\quad}  &\multicolumn{3}{c}{\quad} &
  \multicolumn{3}{c}{\quad} & 4510.0 & 100 & (10) \\
Ladder Creek   & L6     &  S3  &  (15) & {\it 4555.5} & 1.1    & (8) & 4510.8  & 1.0 & (6) &   \multicolumn{3}{c}{\quad} &   \multicolumn{3}{c}{\quad}  & $\surd$  \\
Marion (Iowa)  & L6     & S4 & (15) & {\it 4555.1}  & 1.2  & (8) & 4511.3  & 0.5 & (2) &   \multicolumn{3}{c}{\quad}  &   \multicolumn{3}{c}{\quad}  & $\surd$ \\
\\
Floyd          & L4     &  S3 & (16) &  {\it 4551.5} & 1.2 & (8) \\
Ausson         & L5     &  S3 & (16) & \multicolumn{3}{c}{\quad}  & 4528.0 & 1.7 & (2) \\
Knyahinya      & L5     & S3 & (15) & \multicolumn{3}{c}{\quad} & 4542.6 & 2.4 & (2) \\
Shaw           & L7     & S1 & (15) & \multicolumn{3}{c}{\quad} & \multicolumn{3}{c}{\quad} &  \multicolumn{3}{c}{\quad} & 4420.0 & 40.0 & (5) \\
ALHA-81023     & L5     &    &  & \multicolumn{3}{c}{\quad} & 4534.87 & 1.67 & (6) \\
ALH-85026      & L6     &    &  & \multicolumn{3}{c}{\quad} & 4505.71 & 0.75 & (6) \\
Modoc          & L6     & S2 & (15) & \multicolumn{3}{c}{\quad} & \multicolumn{3}{c}{\quad} & 4509$^{b}$ & 4 & (9) \\
Walters        & L6     & S4 & (16) & \multicolumn{3}{c}{\quad} & \multicolumn{3}{c}{\quad} & 4512$^{b}$ & 4 & (9) \\
Park           & L6     & S1 & (15) & \multicolumn{3}{c}{\quad} &  \multicolumn{3}{c}{\quad} & \multicolumn{3}{c}{\quad} & 4525.8 & 4.6 & (13) \\ 
Kusnashak      & L6     &    &  & 4558.6 & 0.6 & (14) \\
Tennasilm      & L4     &    &  & 4563.5 & 0.7 & (14) \\
NWA6630        & L5     &    &  & 4561.6 & 0.8 & (14) \\
\noalign{\smallskip}
\hline
\noalign{\smallskip}
CAI formation time & & & & 4567.3 & 0.4 & (7) \\
\hline
\end{tabular*}
}

\smallskip\noindent{\scriptsize Notes: (a) feldspar, (b) phosphates, (c) whole-rock data. (d) Without correction for miscalibration of K decay constant \citep{Ren11,Sch11,Sch12}. For this, 30 Ma has be added to each of the data, except for L6 Park.
\\
References: (1) \citet{Spr10}, (2) \citet{Goe94}, (3) \citet{Ame01}, (4) \citet{Rot01}, (5) \citet{Tur78}, (6) \citet{Bla17}, (7) \citet{Con12}, (8) \citet{Spr11}, (9) \citet{Braz99}, (10) \citet{Hut88}, (11) \citet{Pra04}, (12) \citet{Pra06}, (13) \citet{Ruz15}, (14)\citet{Hel19} and personal communication,  (15) \citet{Bis18}, (16) \citet{Sto91}, (17) \citet{Rub94}, (18) \citet{Fri04}. }

\label{TabThermDat}
\end{table*}
 
For comparison with theoretical models we searched the literature on L chondrites for published data on closure times for diffusion of decay products of radioactive nuclei out of the carrier phases. The number of useful data is small because most of the L chondrites are heavily shocked by a catastrophic impact event about 470 Ma ago which likely disrupted the parent body and is thought to be responsible for the high concentration of fossil meteorites in mid-Ordovician marine limestone in southern Sweden \citep{Hey67,Hec04,Kor07,Hec08}. For most of the L chondrites the corresponding data are reset by this event and not useful for our purposes. What one needs are meteorites with low shock grade for which accurate closure temperature data have been determined for at least two different decay systems. One is required for fixing the burial depth of the meteorite in the parent body, a second one (and ideally further ones) is required to fix the properties of the parent body. Data which satisfy this requirement are found for only five meteorites. These are just sufficient to determine the radius and formation time of the L chondrite parent body. For fourteen further meteorites only data of insufficient accuracy or only for one decay system are found. 

In view of the meager data set we construct a somewhat simplified thermal evolution model, which only considers thermal conductivity with temperature dependent heat conductivity and heat capacity, and sintering of the initially porous material. The only free parameters in our model are the radius and the formation time of the body; all other parameters are set to values known otherwise or to plausible values. Additionally we allow for the presence of a regolith layer at the surface with strongly reduced heat conductivity due to impact induced micro-porosity. Because no theoretical model exists for predicting the thickness of such a layer, this is treated as a free parameter to be determined by the model fit. We fit such models to the set of high quality thermochronological data by means of  a genetic evolution algorithm \citep{Cha95}.  We compare the best models found with the data for the individual meteorites and show that it is possible to find a thermal evolution model reproducing the empirical cooling histories of all the meteorites used here.        

Different from the case of H chondrites, two different models with different radii but similar formation times are found to reproduce the empirical data reasonably well, which is not unusual for non-linear optimisation problems. For the smaller one the central temperature stays well below the eutectic melting temperature of the Ni,Fe-FeS system, which is in accord with the fact that no primitive achondrite is known to be compositionally related to L chondrites and may be derived from the same parent body. For the model with bigger radius a small part of the volume shows temperatures above the eutectic melting, but not to such extent, that also differentiation can be expected. This opens the possibility that primitive achondrites from the parent body of the L chondrites might exist, though for some reason none is represented in our meteorite collections. 

\begin{table*}[t]
\caption{Cooling rates (in K\,Ma$^{-1}$) determined for L chondrites. Data of meteorites marked with a hook are used for model optimisation.}

\begin{tabular}{lllclccl}
\hline
\hline
\noalign{\smallskip}
          & Type & Meth.\tablefootmark{1} & Temp. & Rate & Time\tablefootmark{2} & Used & Ref. \\
Meteorite &      &       & [K ]  & [K\,Ma$^{-1}$] & Ma & & \\
\noalign{\smallskip}
\hline
\noalign{\smallskip}
Ausson   & L5 & Pu-ft & 560--370 & 2.2 (+1.3,-0.7) & & & (2) \\
Shaw     & L7 & Pu-ft & 600--365 & 1.9             & & & (2) \\
Mez\"o-Madaras & L3 & met &  773 & 2               & & & (4) \\   
Adelie Land & L5 & met & 773     & 2               & & & (4) \\
Ausson   & L5 & met   &  773     & 1--100          & & & (1), (4) \\
Elenovka & L5 & met   &  773     & 4               & & $\surd$ & (3), (4) \\
Bruderheim & L6 & met &  773     & 12              & & $\surd$ & (4) \\
Holbrook & L6 & met   &  773     & 100--200        & & & (4) \\
Kandahar & L6 & met   &  773     & 6               & & & (3), (4) \\
Mocs     & L5-6 & met &  773     & 20              & & & (3), (4) \\
Tillaberi & L6 & met  &  773     & 400             & & & (4) \\
Waconda  & L6 & met   &  773     & 4               & & & (3), (4) \\
Shaw     & L7 & met   &  773     & $10^3$--$10^4$  & & & (4) \\
Barwell  & L5 & W-diff & $\sim1100$ & $22\pm10$       & $7.6\pm0.7$    & $\surd$ & (5) \\
Bruderheim & L6 & W-diff & $\sim1100$ & $14\pm6$        & $10.4\pm0.8$ & $\surd$ & (5) \\ 
\noalign{\smallskip}
\hline
\end{tabular}
\tablefoot{
\tablefoottext{1}{Pu-ft: plutonium fission tracks, met: metallographic cooling rates, W-diff: W diffusion between silicates and metal.}
\tablefoottext{2}{Time after CAI formation.}
} 
\tablebib{
(1)~\citet{Woo67}; (2) \citet{Pel81}; (3) \citet{Tay71}; (4) \citet{Wil82};
(5) \citet{Hel19}
}

\label{TabMetallDat}
\end{table*}

We also searched the literature on L chondrites for published data on cooling rates derived from Ni concentrations in taenite, from Pu fission tracks, and from W diffusion between silicates and metal by the new approach of \citet{Hel19}. Such data are found for fourteen meteorites, most of them metallographic cooling rates, but all of them are of insufficient accuracy to be useful for the reconstruction of the parent body. They can only be used to check if they are consistent with the constructed model. Despite of this limitation this is of interest because it is claimed in \citet{Tay87} \citep[and emphasized in][ for the case of H chondrites]{Sco14} that metallographic cooling rates are in contradiction to the onion shell model according to which chondrites of petrologic types L3 to L6 originate from a sequence of layers of increasing temperature from the surface to the interior. We find here, as far as this can be judged on the basis of the few available data, that within accuracy limits of the data no discrepancy between empirical cooling rates and those found for the parent body model exists.

The plan of our paper is as follows: Section~\ref{SectData} presents the available meteoritic data, Sect.~\ref{SectModl} describes how the evolution model of the parent body is calculated and Sect.~\ref{SectModFit} how the models are fitted to the empirical data. Section~\ref{SectFinMod} presents some properties of our preferred model for the parent body of the L chondrites, Sect.~\ref{SectComp} compares the model to previous published models, and Sect.~\ref{SectConclu} gives some final remarks. The appendix \ref{AppMatProp} describes details how the required material properties of chondritic matter are determined.  


\section{Data for chondrites}

\label{SectData}

\subsection{Thermochronological data}

Table \ref{TabThermDat} summarizes isotopic age data for L chondrites that are $>4.4$ Ga old and could principally be related to parent body cooling during the first 200 Ma of solar system history. To constrain a cooling path of a certain meteorite within its parent body, at least two data points are needed, i.e., two ages of isotopic systems with distinct closure temperature. These meteorites are listed in the upper part of Table \ref{TabThermDat}: Bruderheim,
Saratov, Elenovka, Homestead, Barwell, Ladder Creek, and Marion(Iowa). Other  meteorites where only one isotopic age data point is available are listed in the lower part of Table \ref{TabThermDat} for completeness, but not used as modeling constraint.

Table \ref{TabThermDat} excludes isotopic ages younger than 4.4 Ga for L chondrites, particularly $^{40}$Ar-$^{39}$Ar ages clustering around 0.47 Ga \citep[see, e.g.,][ and references therein]{Kor07}. These young ages are frequently associated with high shock stages and interpreted as major impact or even impact disruption of the L chondrite parent body \citep[e.g., ][]{Hey67}. Such a late reset of radiometric clocks primarily affects isotopic systems with low closure temperature such as U,Th-He or K-Ar chronometers, and may also
disturb -- though not reset -- chronometers with higher closure temperature such as U-Pb. On the other hand some rocks may have preserved their original isotopic memory by escaping thermal effects of late impacts although some shock effects are discernible in the rocks. Examples of such rocks could be L6 chondrites Bruderheim and Marion (Iowa) in Table \ref{TabThermDat} for which some studies ascribed shock stage S4. An indication of preservation of the earliest isotopic memory is the presence of decay products of short-lived nuclides like $^{182}$Hf and $^{129}$I. While Hf-W ages can be used for cooling curve reconstructions, I-Xe ages are hardly suitable, as for the meteorites in upper part of Table \ref{TabThermDat} (Bruderheim, Saratov, Elenovka) I-Xe ages were obtained on whole rock samples, where the iodine carrier phases and closure temperature is unknown. However, the fact that I-Xe ages could be obtained at least indicates that possible late disturbances did not erase the isotopic memory of in situ decay of $^{129}$I during the first tens of million of years of parent body evolution.

All in all, most of the isotopic age data used for reconstruction of the early parent body history are Hf-W and U-Pb ages with rather high closure temperature, i.e., isotopic systems that are least susceptible to impact related disturbance.

\subsection{Cooling rates}

Reconstructed cooling curves obtained by modeling can further be compared with cooling rates listed in Table \ref{TabMetallDat}. Cooling rate data could be found for twelve meteorites, but only for Bruderheim, Barwell, and Elenovka also thermochronological data are available. Only these can be used for comparison, the other  meteorites with only cooling rate data are listed in Table \ref{TabMetallDat} for completeness, but not used for comparison with models.


\section{Thermal evolution model}

\label{SectModl}

We aim to calculate the temperature evolution within the parent body of the L chondrites and to compare this with the thermochronological data of the set of meteorites discussed in the preceding section. Because we have only relatively few data at hand to compare with, we construct a simple model which depends only on a small set of parameters which we attempt to fix by comparison with the observed properties of the L chondrites. 

\subsection{Basic assumptions}

The basic assumptions are the same as in almost all model calculations on the internal evolution of parent bodies of ordinary chondrites \citep[cf. the review of ][]{McS03}: 

1. The thermal evolution of the body is dominated by the decay of short- and long-lived radioactive nuclei, in particular $^{26\!}$Al and $^{40}$K, and others. Other possible heat sources, in particular external heating, are assumed to have been not important. 

2. The body is formed within a period short compared to the decay time of $^{26\!}$Al from the mixture of dust and chondrules in the solar nebula such that the details of the short initial formation period are not important with respect to the long-term evolution of the body which is the main subject of the model calculation. In the case of the H chondrite parent body such a rapid accretion was found in \citet{Hen13} and \citet{Mon13} to provide the best fit to the meteoritic record. Because the parent bodies of L and H chondrites likely formed in close-by zones of the solar nebula, growth conditions should be similar for both bodies. Then we can start the model calculation with an initially cold body which has already achieved its final radius (instantaneous formation approximation).

3. We assume that the parent body of the L chondrites has a spherically symmetric structure such that the temperature and all other physical variables depend only on the radial distance from the centre, $r$, and on time $t$. 

4. The initial composition of the matter is assumed to be homogeneous within the body. Also the heat production rate per unit mass is assumed to be initially homogeneously distributed within the body and stays to be so during the whole evolution (no differentiation). 

5. We do not consider melting since presently there are no indications that the parent body of the L chondrites was molten and possibly differentiated in its interior. No achondrites or at least primitive achondrites are presently known that can definitely be related to L chondrites, and the L7 chondrites showing evidence of melting are interpreted as impact-melts \citep{Mit01}. Recently, however, a primitive achondrite with oxygen isotopic ratios typical for the L group was found \citep{Vac17} which may indicate that after all there was incipient melting deep in the interior of the parent body of the L chondrites.
It remains to be shown, however, that this meteorite is really related to the same parent body as the other L chondrites.

5. We also do not consider giant impacts by other planetesimals which may disrupt the body already during the early period of its evolution because we aim to study whether it is possible to explain the meteoritical record with respect to temperature evolution with a relatively undisturbed evolution of the parent body over the initial period of heating and cooling during the first about 100 Ma of its evolution. This is not unlikely since according to dynamical calculations 100 km sized bodies have high likelyhood to survive the first 100 Ma un-disrupted \citep{Dav13}. 

\subsection{Equations}

With our assumptions the temporal evolution of $T(r,t)$ is governed by the heat conduction equation 
\begin{equation}
\varrho_\mathrm{b}c_p {\partial\,T\over\partial\,t}={1\over r^2}\ {\partial\over r}\ r^2K\ {\partial\,T\over\partial\,r}+\varrho_\mathrm{b} h\,,
\label{HeatCond}
\end{equation}
where $\varrho_\mathrm{b}$ is the mass density of the asteroid material, $c_p$ its specific heat capacity per unit mass, $T$ the temperature, $K$ the heat conduction coefficient, and $h$ the rate of heat production per unit mass. The quantities $\varrho_\mathrm{b}$, $c_p$, $K$, and $h$ depend on the properties and composition of the material from which the body is built, and in part on temperature and porosity. The dependence of the heat conductivity $K$ on $\Phi$ and $T$ and of the heat capacity $c_p$ on $T$ is complex, as found by  experimental determinations of $K$ and $c_p$ for chondrites  \citep{Yom83, Ope10,Ope12} and in the theoretical investigation by \citet{Hen16} and \cite{Gai18}. The details how the properties of the material are calculated are given in Appendix~\ref{AppMatProp}.

Since the material in ordinary chondrites is subject to heating up to temperatures above 1\,000 K, the composition and physical properties of the mineral mixture change significantly. Besides the temperature dependencies of $K$ and $c_p$, the most important change is reduction of the initial pore space (sintering) since this results in particular in a considerable reduction of the heat conductivity, $K$, and, to a lesser extent, to an increase of the bulk density, $\varrho_\mathrm{b}$, of the mineral mixture. The bulk density, the intrinsic density $\varrho_\mathrm{i}$ of completely compacted material, and the volume fraction of voids, $\Phi$, are related by
\begin{equation}
\varrho_\mathrm{b}=\left(1-\Phi\right)\varrho_\mathrm{i}\,.
\end{equation}

The variation of the porosity depends on temperature, $T$, and the applied pressure, $p$. Theories for calculating the change of $\Phi$ were developed for technical purposes in material science where a set of equations is derived for calculating
\begin{equation}
{\partial\,\Phi\over\partial t}=F\left(\Phi,T,p\right)\,.
\label{EqSint}
\end{equation}
Here we apply the sinter model of \citet{Hel85}. The application of this model to parent bodies of ordinary chondrites is discussed in \citet{Gai15}; it requires to solve a differential equation for $\Phi$ simultaneously with the heat conduction equation.

The pressure $p$ required for calculating the sintering process is determined by the hydrostatic equilibrium equation
\begin{align}
{\partial\,p\over\partial\,r}&=-{GM_r\over r^2}\varrho_\mathrm{b} \\
{\partial\,M_r\over\partial\,r}&=4\pi r^2\varrho_\mathrm{b}\,.
\label{EqMr}
\end{align} 
This equation has also to be solved simultaneously with the heat conduction equation and the equation for sintering. 

\subsection{Initial and boundary conditions}

The heat conduction equation is subject to initial and boundary conditions. Within the instantaneous formation approximation the initial condition is 
\begin{equation}
T(r,0)=T_0\quad(0\le r\le R)\,,
\label{HeatIni}
\end{equation}
where $R$ is the radius of the body and $T_0$ the initial temperature at $t=0$. At the centre of the body the solution has to satisfy the boundary condition
\begin{equation}
{\partial\,T(0,t)\over\partial\,r}=0\quad(t>0)
\end{equation}
to exclude the presence of a point source for heat at the centre. At $r=R$ we  would have to determine the surface temperature from the equilibrium between heating and cooling processes at the surface \citep[see, e.g.,][ ]{Hen12a}. In practice, insufficient information is available for this and we only apply 
\begin{equation}
T(R,t)=T_\mathrm{s}(t)\quad(t>0)
\label{HeatSurf}
\end{equation}
with given surface temperature $T_\mathrm{s}(t)$ as outer boundary .   

In view of the uncertain initial conditions at the time of planetesimal formation and the uncertainties with respect to the environmental conditions experienced by the body during its later evolution, we can only prescribe some plausible value $T_\mathrm{s}$ for the surface temperature and a spatially constant initial temperature $T_0$. A choice for $T_0$ and $T_\mathrm{s}$ has only to obey two conditions: First, the initial temperature and, at least as long as the accretion disk still exists, also the boundary temperature has to be higher than $\sim150$ K because otherwise the parent body of the L chondrites would be formed with abundant water ice which clearly was not the case. Second, the boundary temperature has to be well below the annealing temperature of $\sim390$ K for Pu fission tracks, because such fission tracks are observed for some L chondrites \citep{Pel81}. We chose in our model calculations $T_0=T_\mathrm{s}=200$~K.

The initial porosity $\Phi_0$ of the chondritic material can be estimated from the properties of the unequilibrated meteorites. This is discussed in Appendix~\ref{SectIniPor}. The estimated value of $\Phi_0=0.25$ for the parent body of the L chondrites is used as initial value for the solution of Eq.~(\ref{EqSint}).

\subsection{Solution of the model equations}

The simplest possibility to obtain a solution of the model equations would be to restrict us to the kind of model considered first by \citet{Miy81} who considered only the heat conduction problem and uses for the material properties $\rho$, $c_p$, and $K$ constant values as derived from measured properties of meteorites at room temperature. For constant coefficients, Eq.~(\ref{HeatCond}) can be solved analytically, but in reality the corresponding quantities are by no means constant which strongly changes the thermal structure. Hence, for obtaining physically reliable models we solve Eqs.~(\ref{HeatCond}) to (\ref{EqMr}) numerically which allows us to consider the dependency of $c_p$ and $K$ on temperature and other physical properties of the material. The heat conduction equation and the sintering equation are solved by an implicite finite difference method. The non-linear character of the equations is accounted for by iteration to self-consistency by a fixed point iteration.


\begin{table}

\caption{Closure times, $t_\mathrm{cl}$, closure temperatures, $T_\mathrm{cl}$, and their corresponding errors $\sigma_\mathrm{T}$ and $\sigma_\mathrm{T}$, respectively, used in the model fit.}

\begin{tabular}{llllrl}
\hline
\hline
\noalign{\smallskip}
Meteorite    &       & \multicolumn{1}{c}{$t_\mathrm{cl}$} & \multicolumn{1}{c}{$\sigma_\mathrm{t}$} & \multicolumn{1}{c}{$T_\mathrm{cl}$} & \multicolumn{1}{c}{$\sigma_\mathrm{T}$} \\
             &       &  \multicolumn{1}{c}{[Ma]}     & \multicolumn{1}{c}{[Ma]}   &  \multicolumn{1}{c}{[K]}  & \multicolumn{1}{c}{[K]}  \\
\noalign{\smallskip}
\hline
\noalign{\smallskip}
Bruderheim   & Hf-W  & 4556.9  & 0.8  & 1148 & 50 \\
             & Pb-Pb & 4514.58 & 0.54 &  720 & 50 \\
Elenovka     & Hf-W  & 4558.8  & 0.8  & 1098 & 50 \\
             & Pb-Pb & 4535.0  & 1.0  &  720 & 50 \\
Barwell      & Hf-W  & 4559.7  & 0.7  & 1098 & 50 \\
             & Pb-Pb & 4538.4  & 0.8  &  720 & 50 \\
Ladder Creek & Hf-W  & 4555.5  & 1.1  & 1148 & 50 \\
             & Pb-Pb & 4510.8  & 1.0  &  720 & 50 \\
Marion       & Hf-W  & 4555.1  & 1.2  & 1148 & 50 \\
             & Pb-Pb & 4511.3  & 0.5  &  720 & 50 \\
\noalign{\smallskip}
\hline
\end{tabular}

\label{TabFitDat}
\end{table}

\section{Model fit}
\label{SectModFit}

\subsection{Parameters}

The thermal evolution model depends on a number of parameters, which in part can be derived from the properties and composition of meteorites, and in part on parameters that are not known a priory since they depend on the unknown formation history and formation site of the parent body of the meteorites. The most important unknown parameters are the radius of the body, $R$, its birthtime, $t_\mathrm{b}$, and the thickness, $d$, of an outer impact-modified layer with reduced heat conductivity. Also the initial temperature, $T_0$, at formation time and the radiative equilibrium temperature, $T_\mathrm{s}$, of the surface at the (unknown) distance from the sun are not known. These five unknown parameters can in principle be fixed by determining a set of parameter values for which the deviation between the empirical information on the cooling history of the meteorites and a thermal evolution model takes a minimum, provided a sufficient number of empirical data is available. This is how we proceeded for the H chondrites \citep{Hen13}. We think it better, however, not to overburden the simple thermal evolution model described above by considering details which have merely a limited influence on the model. This holds in particular for the initial and boundary temperatures, T$_0$ and $T_\mathrm{s}$, which are therefore not included in the optimisation but instead are fixed to typical values given in Table~\ref{TabParRange}. Therefore, we restrict the set of model parameters to the birthtime, $t_\mathrm{b}$, the radius, $R$, of the parent body, and the thickness of the insulating layer, $d$. This set of parameters is considered as the best information on the likely properties of the parent body of the L chondrites that can presently be obtained. 

\begin{table}
\caption{Range of parameter values that were allowed in the optimisation process, and the initial values.}

\begin{tabular}{llll}
\hline
\hline
\noalign{\smallskip}
\noalign{\smallskip}
Parameter & \hspace{-.3cm}Symbol & Range & Unit \\
\hline
\noalign{\smallskip}            
                    & \multicolumn{3}{l}{\it Parameter range} \\[.1cm]
Birthtime           & $t_{\rm b}$ & 1.5 -- 2.5 & Ma \\
Radius              & $R$            &  50 -- 300 & km \\
Insulating layer    & $d$            & 0 -- 9     & km  \\[.2cm]
                    & \multicolumn{3}{l}{\it Initial values} \\[.1cm]
Surface temperature & $T_{\rm s}$    & 200        & K \\
Initial temperature & $T_0$          & 200        & K \\
Initial Porosity    & $\Phi_0$       & 0.25       &  \\
\noalign{\smallskip}
\hline
\end{tabular}

\label{TabParRange}
\end{table}

\subsection{Comparison with meteoritic data}

The data set for L chondrites to be compared with evolution models for their parent body is summarised in Section~\ref{SectData}. We can use only the seven L chondrites Bruderheim, Saratov, Elenovka, Homestead, Barwell, Ladder Creek, and Marion (Iowa), for which useful data on closure ages for at least two different thermochronometers with different closure temperatures are available. In the fitting procedure between models and meteoritic data we omit, however, the data for Saratov and Homestead because their Hf-W ages may not have been completely reset by mild metamorphism and partly keep the memory of chondrule formation processes (J. Hellmann, personal communication).  The final data set used for model fitting is shown in Table \ref{TabFitDat}; data are much more limited than for H chondrites, but should in principle suffice to perform a meaningful fit. 

\begin{figure*}

\includegraphics[width=.6\hsize]{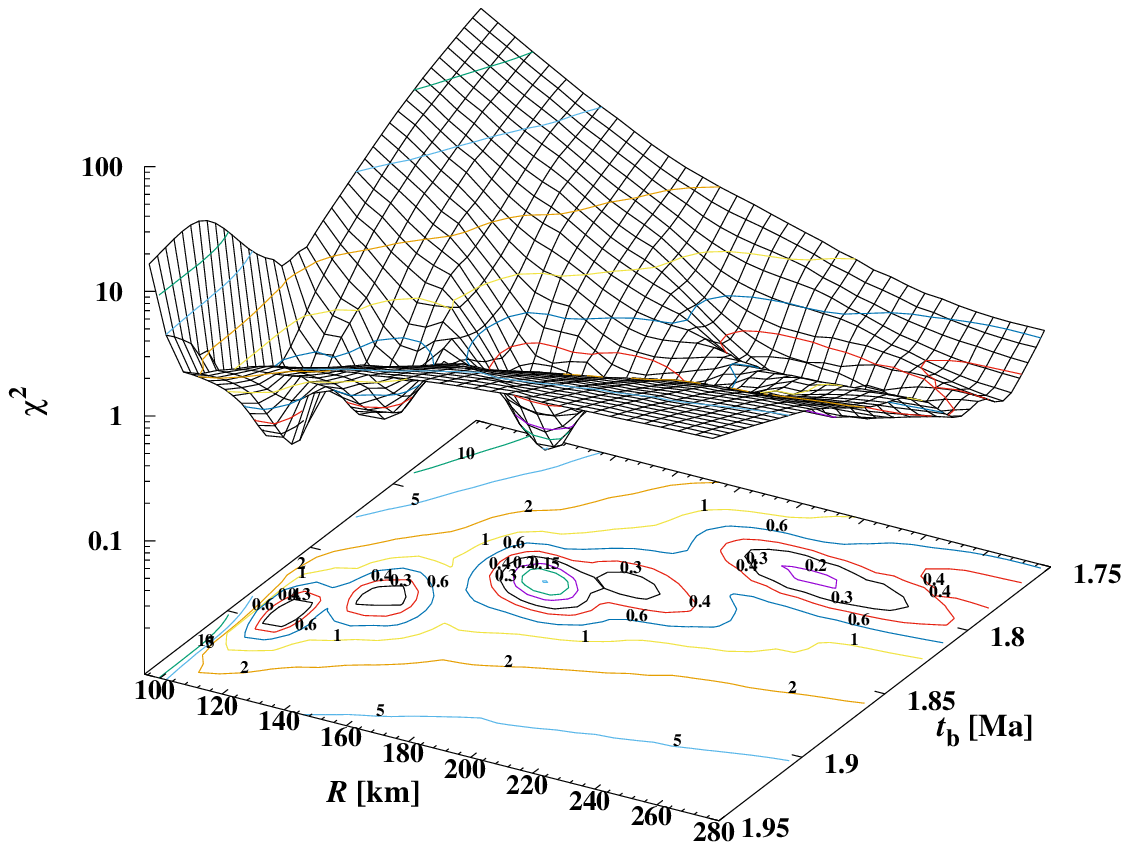}
\hfill
\includegraphics[width=.39\hsize]{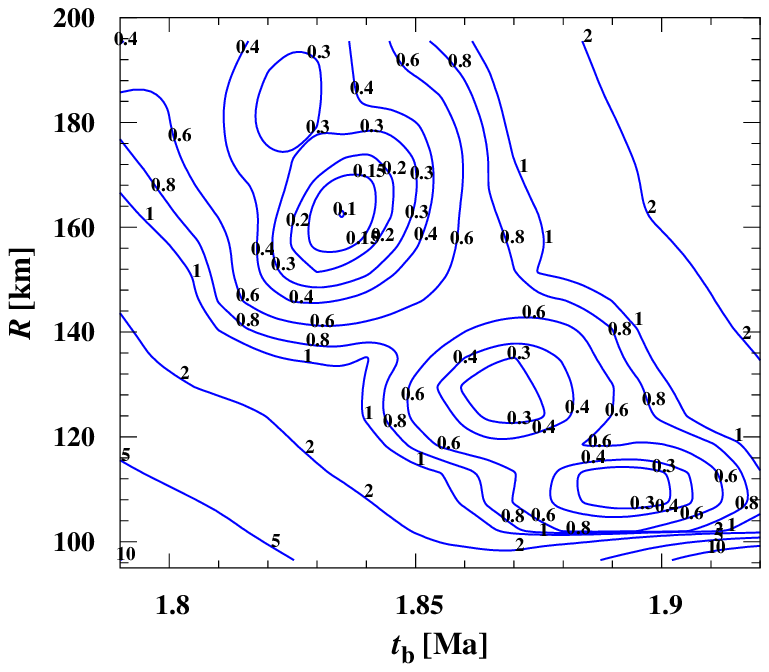}

\caption{Variation of $\chi^2$ with radius $R$ of the body and birthtime $t_\mathrm{b}$ for fixed thickness $d$ of the outer impact-damaged layer. \emph{Left part:} Shape of the $\chi^2$-surface. Also shown are contour lines of constant values of $\chi^2$ on the $\chi^2$-surface and their projection onto the basis-plane. \emph{Right part:} Contour lines of $\chi^2$ in the $R$-$t_\mathrm{b}$-plane.
}

\label{FigSurfChi2}
\end{figure*}

In the following $m$ enumerates the meteorites used. For each of these we know for totally $N(m)$ different thermochronologic systems closure temperatures $T_{m,n}$ and closure times $t_{m,n}$ and associated errors $\sigma_{T,m,n}$ of the closure temperature, where $n$ enumerates these systems. The burial depths of the meteorites, $d_m$, are determined by first determining for a given model and a given meteorite that depth for which 
\begin{equation}
\chi^2_m=\sum\limits_{n=1}^{N(m)}
{\left(T(t_{m,n})-T_{m,n}\right)^2
 \over
\sigma^2_{T,m,n}
}\,.
\end{equation}
takes a minimum. Here $T(t_{m,n})$ is the model temperature at the burial depth of the metorite. The minimum is determined by calculating for some grid of burial depths the values of $\chi^2_m$ and determining the minimum from fitting a parabola through the lowest value and that of the two adjacent points of the grid. The location at which the minimum is taken corresponds to the burial depth of the meteorite. The quality function then is 
\begin{equation}
\chi^2=\sum_m\chi^2_m+{\cal P}
\end{equation}
with the values of $\chi^2_m$ as found in the determination of the burial depths.

We can add a penalty function ${\cal P}$ to $\chi^2$ to enforce the model to obey additional restrictions, e.g., to guarantee that the central temperature stays below the eutecting melting temperature of the Ni,Fe-FeS system. In this latter  case we could chose $\cal P$, e.g., as follows
\begin{equation}
{\cal P}=\max\left(T_\mathrm{c}-T_\mathrm{m},0\right)\,,
\label{DefPenTc}
\end{equation}
where $T_\mathrm{c}$ is the maximum central temperature of the model and $T_\mathrm{m}$ the melting temperature.

For obtaining a best fit between data and model results,~i.e., for  minimizing the quality function, $\chi^2$, a fitting routine is needed for which we adopt the genetic algorithm PIKAIA of \citet{Cha95} which is, as any genetic algorithm, in principle able to find the global minimum even for highly complicated problems. This has also been used in our preceding papers and found to be very useful for this purpose.

In our calculation we vary three model parameters to optimize a model, the radius, $R$, the birth time, $t_\mathrm{b}$ of the body, and the thickness, $d$, of an impact-damaged surface layer. The parameters are varied each within a range of physically meaningful values which are given in Table \ref{TabParRange}. This limitation of the range of variation serves to prevent the algorithm from trying to explore unphysical regions of the parameter space.

For an optimisation we ran the genetic code for at least 40 generations with 50 individuals and we performed always at least ten optimisations with different seeds for the random number generator in order not to run into the risk that the algorithm gets stuck for a large number of steps in a local minimum and the calculation is stopped too early before a mutation pushes the algorithm to continue the search for a better result in another region of the parameter space. We used a resolution of the parameter values of four digits \citep[see][for the meaning of this]{Cha95}.

\begin{table}
\caption{Resulting model parameters, central temperature, and fit quality of three possible models (called LM1, LM2, and LM3) for the L chondrite parent body.}

\begin{tabular}{llllll}
\hline
\hline
\noalign{\smallskip}
\noalign{\smallskip}
Parameter & \hspace{-.3cm}Symbol & LM1 & LM2 & LM3 & Unit \\
\hline
\noalign{\smallskip}
Birthtime           & $t_{\rm b}$ & 1.884  & 1.888  & 1.835  & Ma \\
Radius              & $R$            & 101.4  & 115.2  & 159.8  & km \\
Insulating layer    & $d$            & 4.46   & 2.91   & 3.51   & km \\
Central \\
Temperature         &                & 1\,183 & 1\,186 & 1\,254 & K  \\
\noalign{\smallskip}
\hline
\noalign{\smallskip}
Fit quality         & $\chi^2$ & 0.166 & 0.153 & 0.0975 & \\
Normalised          & $\chi^2_{\rm n}$ & 0.0832 & 0.0763 & 0.0488 & \\
\noalign{\smallskip}
\hline
\end{tabular}

\label{TabParVar}
\end{table}

\subsection{Optimised model}

Ten optimisations with different seeds of the random number generator were run for the range of reasonable parameter values, as defined in Table \ref{TabParRange}, with 40 generations and 50 individuals each. The optimisation resulting in the lowest value of $\chi^2$ (usually there are more than one with almost equal value of $\chi^2$) was re-run with 100 generations. The optimum model was found in step 1\,884 of 5\,000 and no improvement was found in the following steps. This suggests that we really have found the optimum model corresponding to the lowest value of $\chi^2$. We found, that within the range of admitted parameter values some additional well defined local minima of the quality function $\chi^2$ exist where $\chi^2$ takes a local minimum value comparable to that of the optimum model. The set of parameter values corresponding to the optimum model and to two of the alternative local minima are shown in Table \ref{TabParVar}. These models are labelled as LM1, LM2, and LM3, with LM3 being the optimum model. If we apply the penalty function (\ref{DefPenTc}), only models LM1 and LM2 are obtained. Models LM1 and LM2 have similar low values of the quality function $\chi^2$ as the optimal model LM3 and therefore fit the empirical data nearly equally well.  Table \ref{TabOptMet} presents the resulting burial depths of the meteorites and the peak temperatures achieved during the evolution at the burial depths. 

Table~\ref{TabParVar} also shows the normalised fit quality
\begin{equation}
\chi^2_{\rm n}=\chi^2/(D-P)
\end{equation}
which is a better indicator for a proper characterisation of the minimum. Here $D$ is the number of data points fitted (in this case $D=10$) and $P$ the number of parameters (which is $P=8$, the three varied model parameters plus the five burial depths that have to be determined). For a resulting value of $\chi^2_{\rm n}<1$ the model can be considered as agreeing with the data sufficiently well such that we can assume that it reproduces the meteorite cooling curves as documented by the thermochronological data, whereas a value of $\chi^2_{\rm n}\gtrsim1$ would mean that there is a significant misfit between model and data. The value $\chi^2_{\rm n}\approx 0.0109$ for the optimum model LM3 given in Table~\ref{TabParVar} satisfies  $\chi^2_{\rm n}\ll1$ which means that for our evolution model a set of parameter values has been found for which the corresponding model reproduces the data. Models LM1 and LM2 also satisfy this condition. 

\begin{table}

\caption{Burial depths $d_\mathrm{b}$ of the L chondrites in the optimized models and maximum temperature $T_\mathrm{max}$ achieved during evolution.}

\begin{tabular}{lcccccc}
\hline
\hline
\noalign{\smallskip}
              & \multicolumn{2}{c}{LM1} & \multicolumn{2}{c}{LM2} & \multicolumn{2}{c}{LM3} \\
              & $d_\mathrm{b}$ & $T_\mathrm{max}$ & $d_\mathrm{b}$ & $T_\mathrm{max}$ & $d_\mathrm{b}$ & $T_\mathrm{max}$ \\
Meteorite     & [km] & [K] & [km] & [K] & [km] & [K] \\
\hline
\noalign{\smallskip}
Bruderheim    & 35.5 & 1159 & 35.9 & 1156 & 28.0 & 1187 \\
Elenovka      & 20.2 & 1127 & 22.0 & 1128 & 18.0 & 1149 \\
Barwell       & 18.1 & 1116 & 20.0 & 1118 & 16.0 & 1134 \\
Ladder Creek  & 41.8 & 1163 & 42.0 & 1160 & 31.2 & 1192 \\
Marion(Iowa)  & 41.7 & 1163 & 41.8 & 1159 & 31.3 & 1193 \\
\noalign{\smallskip}
\hline
\end{tabular}

\label{TabOptMet}
\end{table}

To get an impression of how well it is possible to fit the set of available thermochronological data of the L chondrites by our thermal evolution model we show in Fig.~\ref{FigSurfChi2}a the $\chi^2(R,t_\mathrm{b})$ surface and the contour lines of constant $\chi^2$ in some range around the optimum set of parameter values for varying $R$ and $t_\mathrm{b}$ and for a fixed value of $d$ equal to its optimum value. We easily recognize the well-defined minimum in the $\chi^2$ surface around the optimum value of $R$ of 160 km and a formation time of 1.84 Ma. Figure~\ref{FigSurfChi2}b shows in a more close-up view the projections of a number of contour lines onto the $R$-$t_\mathrm{b}$-plane around the position of the optimum $R$-$t_\mathrm{b}$-values at fixed $d$. An inspection shows that the formation time and the radius of the body are rather well determined if $d$ is held fixed at its optimum. A deviation of $\chi^2$ by a factor of two from its optimum corresponds to a deviation of the formation time from its optimum value by about $\pm0.01$ Ma, and a deviation of the radius from its optimum  value by about $\pm10$ km. This indicates the accuracy with which the formation time and the radius of the parent body can be determined in our case. 

\subsection{Alternative models}

Additionally one finds minima of $\chi^2$ at radius $>200$ km and at $R\lesssim140$~km, but with a somewhat bigger value of $\chi^2$ at minimum. The bigger body would be of the size of Vesta and as such is unlikely to have been completely disrupted during the lifetime of the solar system. Since Vesta is not the source of the L chondrites and no other body (except for Ceres) of this size is found in the Asteroid belt, the bigger model is rejected. 

\begin{figure}
\includegraphics[width=\hsize]{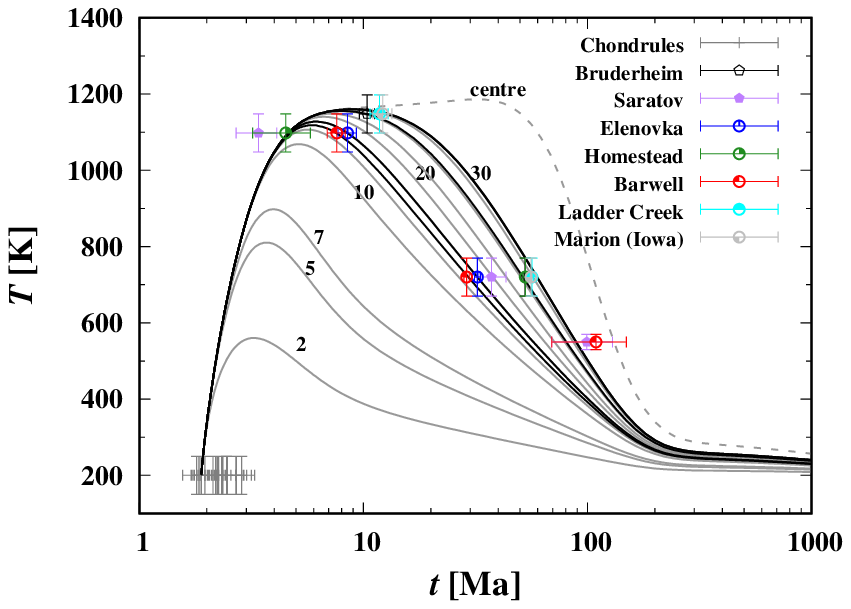}

\includegraphics[width=\hsize]{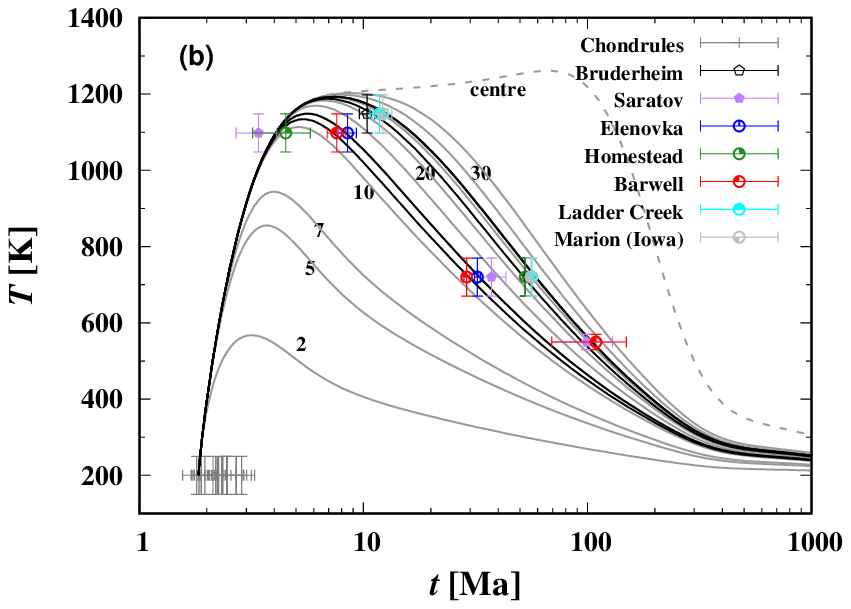}

\caption{Temperature evolution at different depths below the surface (grey lines) and evolution of the central temperature (dashed grey line). The symbols with error bars correspond to the closure temperatures and closure times of L chondrites. The black lines correspond to the temperature evolution at the burial depths of the five meteorites from Table~\ref{TabOptMet} for (a) model LM2 and (b) model LM3. The data of Homestead and Saratov are not used for the optimisation. Grey dots show chondrule ages for L chondrites from \citet{Pap19}.
}

\label{FigFitMet}
\end{figure}

\begin{figure*}

\includegraphics[width=0.32\hsize]{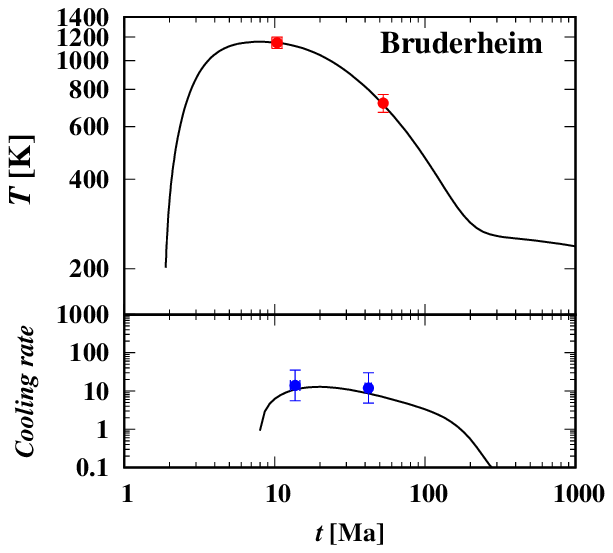}
\includegraphics[width=0.32\hsize]{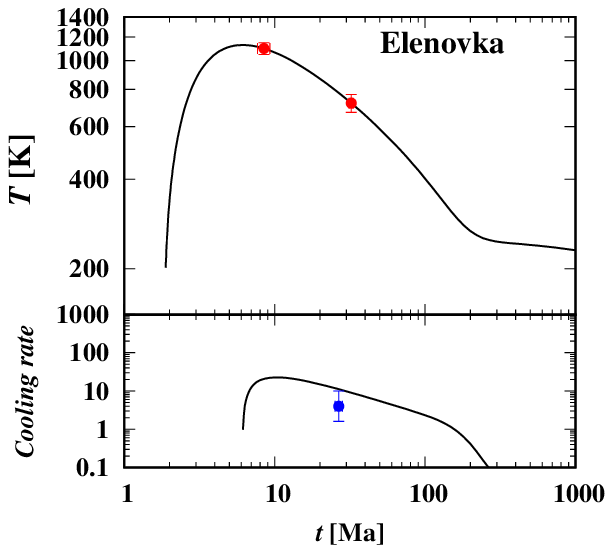}
\includegraphics[width=0.32\hsize]{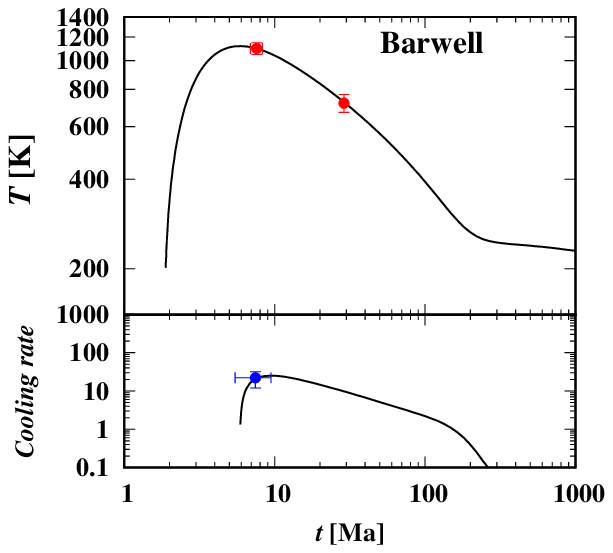}

\includegraphics[width=0.32\hsize]{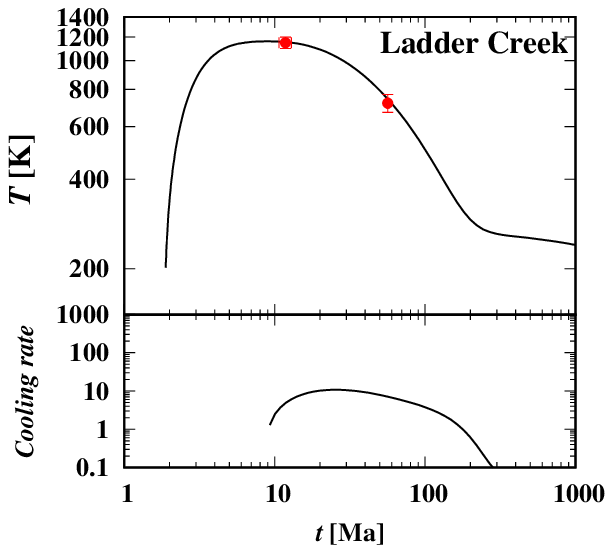}
\includegraphics[width=0.32\hsize]{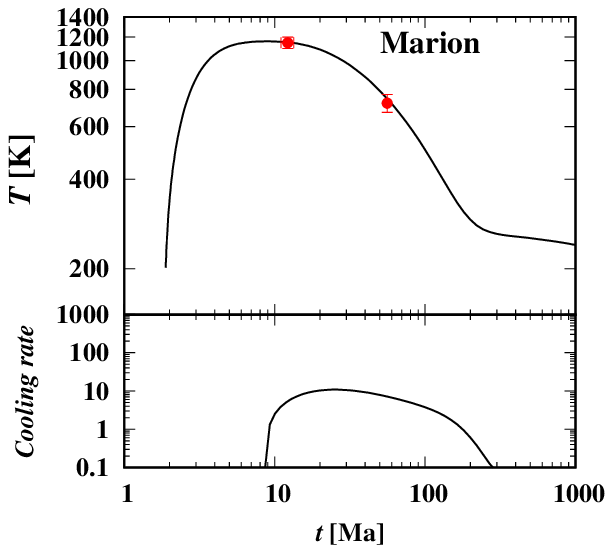}

\caption{Temperature evolution at the burial depths of the meteorites and the corresponding cooling rates (in K per Ma). The dots with error bars show the empirical data. Red dots show the data used for the optimisation, blue dots cooling rates. 
}

\label{FigComMetMod}
\end{figure*}
  
For the two other local minima in the $\chi^2$ surface at $R\lesssim140$~km we performed an optimisation with a restricted search volume in parameter space. The resulting optimum parameter values and the corresponding values of $\chi^2$ are given in Table~\ref{TabParVar} as models LM1 and LM2. In both cases $\chi^2_\mathrm{n}$ is quite small, though not as small as for model LM3, but both models also fit the empirical data quite well. For models LM1 and LM2 there is an important difference in the model properties compared to LM3: The  central temperature definitely stays below the temperature for eutectic melting of the Ni,Fe-FeS system of 1\,223~K \citep{Mar14}, while for model LM3 it definitely exceeds this melting temperature. 

If models LM3 should correctly describe the properties and early evolution of the L chondrite parent body, then there could exist primitive achondrites with signatures of incipient eutectic melting of Ni,Fe-FeS which show a clear relation to L chondrites according to their composition. If, on the other hand, no such primitive achondrites related to L chondrites are found, models LM1 or LM2 probably describe the properties and early evolution of L chondrites. In fact, despite of the L chondrites being the most abundant known ordinary chondrites with presently about 7100 known specimens, no primitive achondrite clearly related to the class of L chondrites has ever been found. This suggests models of this type to be the appropriate one. Because of its slightly better fit quality, we prefer in the following model LM2.

Recently, however, a primitive achondrite with oxygen isotopic ratios typical for the L group was found \citep{Vac17} which may indicate that after all there was incipient melting deep in the interior of the parent body of the L chondrites. This would point to model LM3 as the correct one, but it remains to be shown that this meteorite is really related to the same parent body as the other L chondrites.

\section{Model results}

\label{SectFinMod}

\subsection{Fit of meteoritic data}

In order to demonstrate how the models fit to the meteoritic data Fig.~\ref{FigFitMet} shows the variation of temperature at a number of selected depths during the course of the evolution for models LM2 and LM3. Also shown are the thermochronological data for the meteorites for which at least data of acceptable accuracy for a minimum of two thermochronological systems are available, including the meteorites that are not used in the optimisation. The thermal evolution of the five meteorites from Table~\ref{TabOptMet} is shown as black lines. The temperature curves for each of these meteorites pass right through the centre of the error boxes for almost all of their thermochronological data (see Fig.~\ref{FigFitMet}). The fit therefore is quite good, which may in part result from the circumstance that we only use those data for the optimisation which can be considered as reliable and highly accurate. If one includes also data of lower accuracy and unclear reliability from Table~\ref{TabThermDat} the fit quality would become much worse. 
  
The quality of the fits to the data is practically indistinguishable for the two kinds of models with either no melting in the core region or with incipient melting in the core. The presently available data do not allow to discriminate between the smaller or bigger sized body on the basis of the thermochronological data. It is only the lack of primitive achondrites related to L chondrites that favours model LM2. 

We now consider the individual fits between empirical data for the cooling history of a meteorite and the thermal evolution model of the parent body for the five meteorites used for the optimisation. Figure~\ref{FigComMetMod} shows in the upper part of the pictures the temperature evolution of the favoured model LM2 at the burial depth of the meteorites. In the lower part of the pictures the cooling rate after culmination of the temperature is shown. In the upper part we also show the data points of the closure-time and closure-temperature of the radioisotopic clocks used for the optimisation as red dots with error bars for closure time and closure temperature. In the lower parts of the picture we show data for cooling rates for the three meteorites for which besides radioisotopic clock data also such kind of data are available. The figure confirms again that it is possible to obtain a very good fit between the highly accurate Hf-W and Pb-Pb data and an evolution model for the parent body. 

Not used for the model fit are cooling rates, if such data exist, because of their generally low accuracy. For the meteorites shown in Fig.~\ref{FigComMetMod}  metallographic cooling rate are published for Elenovka and Bruderheim, and cooling rates determined by the new method of \citet{Hel19} for the Hf-W system are available for Barwell and Bruderheim. 

In case of the metallographic cooling rates we derive for a closure temperature of 773~K for Ni diffusion in taenite from the temperature run at the known burial depth a closure time and plot the corresponding data point in the lower part of the pictures. The accuracy of the cooling rates derived from Ni diffusion in taenite is estimated by \citet{Woo67} as of about a factor of 2.5.  

For the Hf-W-sytem the time to which the cooling rate refers is given directly in \citet{Hel19}. 

\begin{figure}

\includegraphics[width=\hsize]{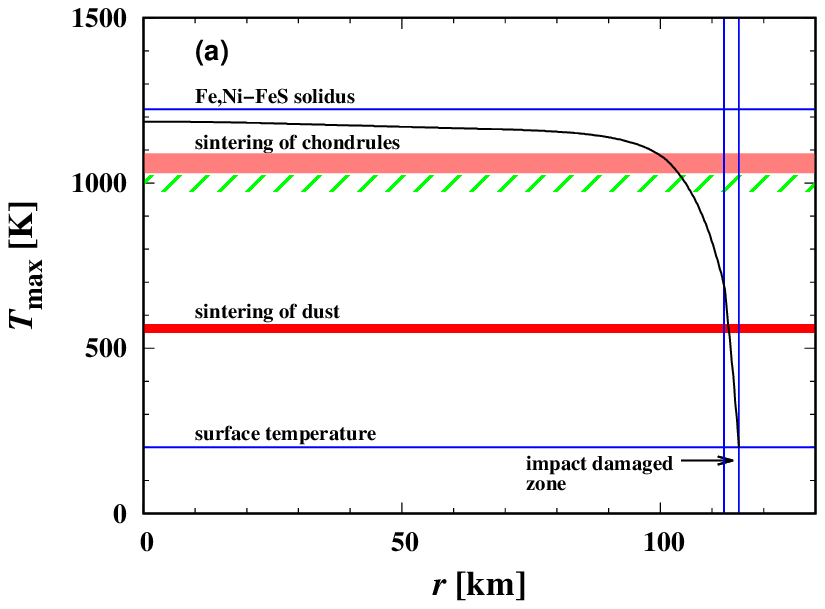}

\includegraphics[width=\hsize]{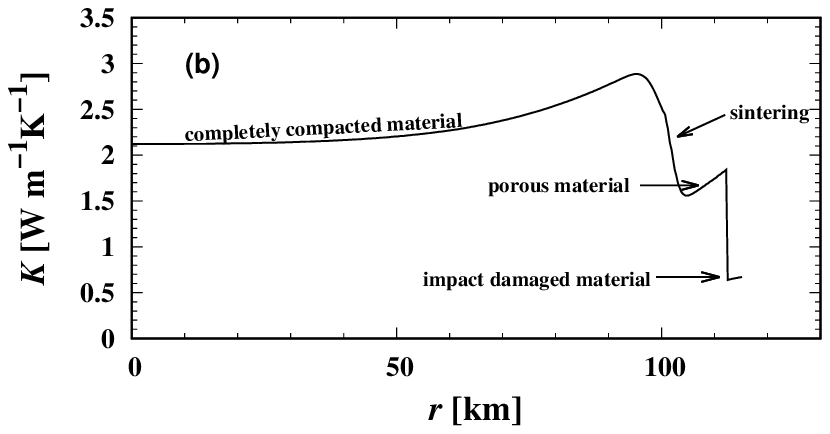}

\caption{(a) Maximum of temperature at radius $r$ during the undisturbed thermal evolution of the body (black line) for model LM2. The two broad red strips indicate the temperature range over which the chondrule assembly (upper strip) and the initially fine grained matrix material (lower strip) are compacted. The hatched area is the temperature range between the upper metamorphic temperatures of L4 and of L5 chondrites  \citep[973 K to 1023 K, respectively, according to][]{McS88} where the intergranular voids between chondrules disappear. The upper blue line shows the eutectic melting temperature of 1\,223 K for the Ni,Fe-FeS system at the Ni content of L chondrites. (b) Radial variation of the heat conductivity $K$ at 30 Ma after formation of the body.
}

\label{FigTmax}
\end{figure}

Totally we have four data for cooling rates for three meteorites which can be compared to our parent body model. For these empirical cooling rates the cooling rate curve at the burial depth of the parent body model agrees within the error limits with the observed cooling rates. This suggests that the cooling rates of the model for the parent body also agree reasonably well with the empirical cooling rates. 

In summary, we conclude that it is possible to find a model that reproduces the observed cooling history of all L chondrites for which presently accurate thermochronological data are available.
 
\begin{figure}

\includegraphics[width=\hsize]{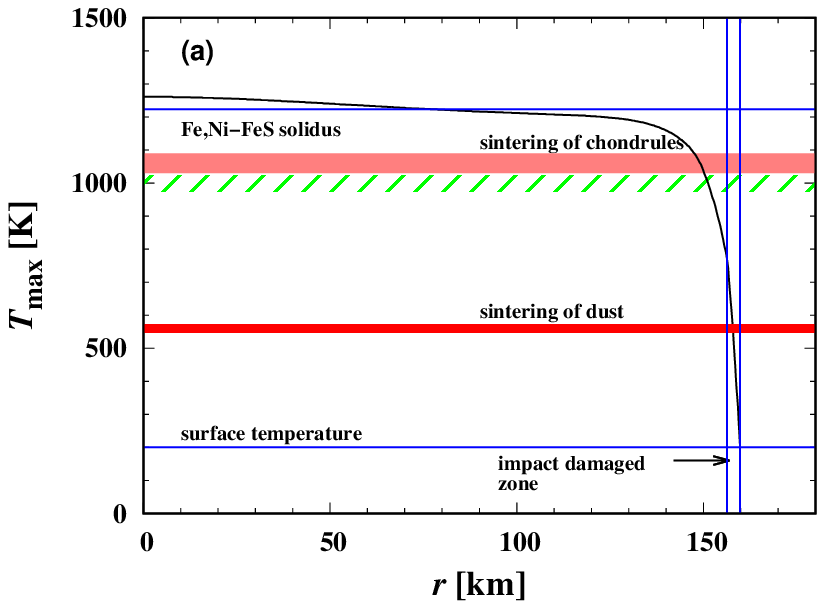}

\includegraphics[width=\hsize]{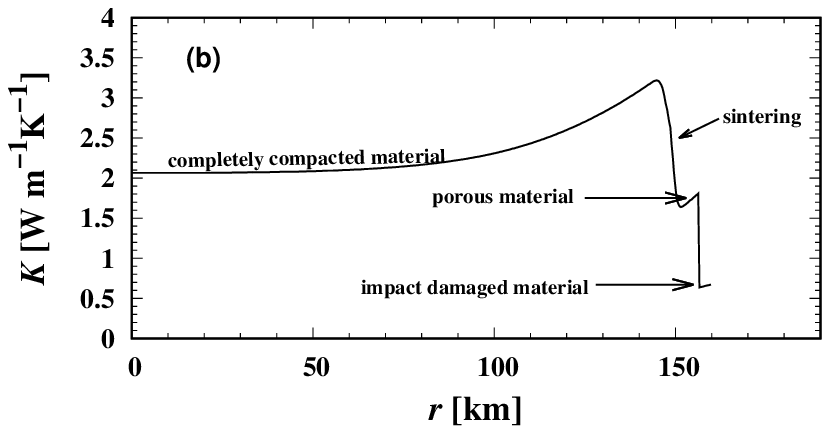}

\caption{As Fig.~\ref{FigTmax}, but for model LM3.}

\label{FigTmaxBig}
\end{figure}

\subsection{Age of chondrules}

As chondrules formed as individual millimeter sized objects in the solar nebula before accretion of the parent asteroid, the parent body formation time resulting from our model serves as lower limit for the age of chondrule formation. Figure \ref{FigFitMet} shows $^{26\!}$Al/$^{26}$Mg ages for a number of individual chondrules from types L3.00 to L3.15 from \citet{Pap19}. These are probably the most accurate chondrule ages presently known. They range from an average age of $1.80^{+0.11}_{-0.10}$ Ma for the oldest chondrules up to $\sim3$ Ma after formation of CAIs. They are plotted in Fig.~\ref{FigFitMet} with error bars corresponding to the possible range of initial temperatures for ordinary chondrite parent bodies. The age of the oldest chondrules is interpreted by \citet{Pap19} as ``to record the major and relatively punctuated onset of chondrule formation". If this interpretation is correct, our result on the birthtime of the parent body of the L chondrites suggests that the formation of the body followed almost immediately the onset of chondrule formation, within $\sim0.03$ Ma if model LM3 is applicable or within $\sim0.08$ Ma, if model LM2 is applicable, i.e., within the errors bounds for the time of onset of chondrule formation.

The ages of the chondrules in Fig.~\ref{FigFitMet} are for chondrules from the most primitive unequilibrated meteorites with the lowest petrologic types L3.00 to L3.15 of ordinary chondrites. They resided in layers within a few km below the surface of the body which most likely acquire some additional later-formed material even after the main formation period of the body has terminated. The spread in chondrule ages over a period of more than 1 Ma, thus, is not necessarily in conflict with the instantaneous formation period hypothesis used in our model calculations but hints either to an extended period of chondrule formation over 1.2 Ma duration \citep{Pap19} or subsequent disturbance of the Al-Mg system. 

\subsection{Properties of the parent body}

In the following we consider some consequences of the thermal evolution of the body. The variation of temperature at fixed radius during the course of the thermal evolution of the body is shown for some selected radii in Fig.~\ref{FigFitMet}. There is a rapid increase in temperature within~$\sim3$ Ma immediately after formation of the body, a period of high temperature lasting several to tens of Mega years, and a final cooling period lasting tens to hundreds of Mega years. 

The evolution of the temperature at the centre of the body is shown by the dashed line. The central temperature takes a maximum at about 40 Ma for model LM2 and 80 Ma for model LM3. The corresponding flat hump in Fig.~\ref{FigFitMet}b is due to the contribution of $^{40}$K to the heating of the body. Without this the temperature inside the body would decrease significantly earlier which would modify also the thermal evolution at the lower depths where the meteorites come from (cf. Fig. \ref{FigHeat26Al40K}).  

An important aspect of the thermal history is the highest temperature, $T_\mathrm{max}$, found at a given radius since this determines the degree of metamorphism from the complex initial mixture of dust and chondrules seen in petrologic type 3 chondrites to the compacted and crystallized rocky material seen in type 6 chondrites, or, may be, even to partially molten material of primitive achondrites. The radial variation of this maximum temperature in the model is shown in Fig.~\ref{FigTmax}a for model LM2 and in Fig.~\ref{FigTmaxBig}a for model LM3.    
 
There are three characteristic temperatures where the structure of the material and its heat conductivity significantly changes. The first is where the initially porous matrix material is compacted by surface diffusion \citep{Gai15}, the  second is where the ensemble of chondrules is compacted by deformation creep \citep{Gai15}, and the third where the first components of the complex mixture of minerals and metal of the chondritic material start melting.

The compaction of matrix and chondrules extends over some temperature range which is indicated in Figs.~\ref{FigTmax}a and \ref{FigTmaxBig}a by two red-coloured strips. Each of the two processes  result in a significant increase of the strength of the material. At about $560\pm20$~K the probably rather loosely bound initial dust + chondrule mixture bakes together into a sandstone-like material. Around $1060\pm30$~K creep processes of the minerals in chondrules become thermally activated such that under the action of pressure inside the body the remaining pore space between chondrules disappears. At the same time at such high a temperature crystal growth of the mineral components and diffusional re-distribution of iron into coarser inter-chondrule-space filling iron grains becomes active, which altogether act to form a rocky material of significant strength.

The overall run of $T_\mathrm{max}(r)$ with distance $r$ from the centre as seen in Figs.~\ref{FigTmax}a and \ref{FigTmaxBig}a is rather flat across most part of the body and steeply drops to the low surface temperature within a thin surface layer. This nearly isothermal interior of the body and the steep gradient near the surface results from efficient heat conductivity in the compacted interior of the body and low heat conductivity in the porous and impact-damaged structure of the surface material. 

The variation of $K$ with radius in our model of the L chondrite parent body is shown in Figs.~\ref{FigTmax}b and \ref{FigTmaxBig}b. Since the heat conductivity depends on temperature and, thus, on time, we show the heat conductivity at 30 Ma and 65 Ma after the formation of the body for models LM2 and LM3, respectively, which is representative for the period where the temperature in the central region is highest. The transition from the porous outer material to the completely compacted material in the interior occurs at a depth between 11.0 and 17.5 km below the surface in model LM2 and between 8.9 and 11.5 km below the surface in model LM3. The second strong change in $K$ occurs at a depth of 2.9~km below the surface in model LM2 and at 3.5~km below the surface in model LM3. This is because we allowed for in our model that there is a surface layer where the material is heavily damaged by impacts and where numerous micro-cracks in the mineral component of the material strongly reduce the heat conductivity to a low value as it is also measured for part of the chondrites (see Fig.~\ref{FigKheat}b).

According to the model optimisation such a layer is required to obtain a good fit to the empirical data. Since we presently cannot predict $d$ from a theory of the impact induced modifications of the properties of the surface material on an asteroid, it is hard to decide, whether our optimum value of $d$ obtained from the fitting process is physically justified  or simply allows to compensate for other deficits of the model. But at least one point seems clear: A reliable fit with an only very thin or even absent insulating surface layer cannot be obtained for the L chondrite parent body. The situation would probably improve if not only data for L5 and L6 chondrites are available, but also data for undisturbed L4 chondrites. They are more sensitive to the temperature at shallower surface layers and would allow to better to constrain the properties of the outer part of the body.  

\subsection{Onion shell model}

It is generally assumed that the metamorphism of the chondritic material from the highly pristine material seen in L3.0 meteorites to the highly equilibrated material seen in L6 meteorites is essentially determined by the peak temperature experienced by the material. According to the onion-shell model hypothesis this thermal metamorphism proceeded in different depth regions of the parent body experiencing different maximum temperatures, as shown in Fig.~\ref{FigTmax}, during the course of the thermal evolution of the body. Meteorites of type L6 correspond in this model to material from the core region, type L3 to material from a layer just below the surface, and types L4 and L5 from layers of increasing depth and temperature below the surface. The basic assumption then is, that the body experienced no disruptive collision within at least the first 100 to 200 Ma of its internal thermal evolution, such that all layers cooled below the closure temperatures of the thermochronological systems and at least some meteorites can be found where their thermochronological clocks remained essentially undisturbed if the body is shattered into a rubble pile.

\begin{figure}

\begin{center}
\includegraphics[width=.8\hsize]{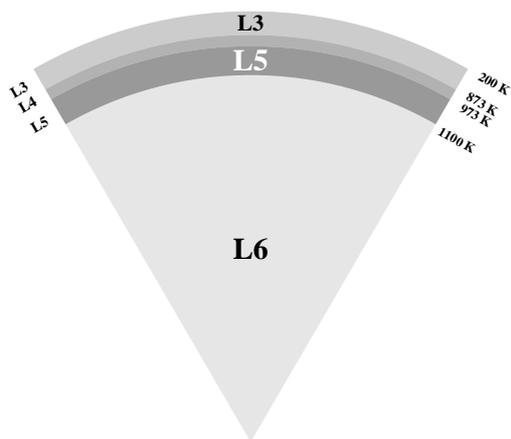}
\end{center}

\caption{Schematic sketch of the onion-shell structure of the L chondrite parent body.
}

\label{FigOnionL}
\end{figure}
\begin{table}

\caption{Maximum temperature of petrologic types 3 \dots\ 6 and corresponding burial depths $b$ where maximum of temperature equals the maximum metamorphic temperature for model LM2.
}

{\small
\begin{tabular}{l@{\hspace{.2cm}}lcccc}
\hline
\noalign{\smallskip}
Border     &      & {L3 - L4} & {L4 - L5} & {L5 - L6} & {L6 - L7} \\
$T_{\max}$ & [K]  & 873  & 973  & 1138 & 1173 \\
Depth $b$  & [km] & 6.35 & 9.31 & 17.5 &      \\ 
\noalign{\smallskip}
\hline
\end{tabular}
}

\label{TabMaxTempPetro}
\end{table}

Since the seminal paper of \citet{Yom83} the boundaries between regions where the material shows characteristics corresponding to the different petrologic types are calculated from maximum metamorphic temperatures of a given petrologic type. We take data for these temperatures for type L3 and L4 (873 K and 973 K, respectively) from \citet{McS88} and for L6 (1173 K) from \citet{Sla05}. The maximum peak temperatures are generally highly uncertain \citep[about $\pm50$ K, e.g. ][]{Kes07}, in particular for petrologic types 4 and 5 for which they are only estimates. Especially the temperature of 1023 K for the border between L5 and L6 as given by \citet{McS88} seems problematic since with this value for the border temperature the width of the temperature range for type L5 between the lower temperature border separating L5 from L4 and the higher temperature border separating L5 from L6 is of the same order of magnitude as the error of the two border temperatures. Therefore we proceed in this case as follows: According to the classification of \citet{vaS67} one major difference between petrologic types 5 and 6 is that for type 6 chondrules are poorly defined, while for type 5 chondrules are readily delineated. We then identify the border between types 5 and 6 by the disappearence of chondrules by sintering as easy-to-recognize individuals and take (somewhat arbitrarily) a residual pore space of the granular material of 5\% as limit, which translates according to our model into a temperature at the border between L5 and L6 of $1\,100\pm25$ K (the error limits corresponds to almost none to almost complete sintering).

For our preferred model LM2 the depths at transitions between the different petrologic types with these values of temperature are given in Table \ref{TabMaxTempPetro}. A schematic sketch of the distribution of petrologic types in the body  is shown in Fig.~\ref{FigOnionL}. The volume fractions of types L3 \dots\ L6 are considered in Sect.~\ref{SectAbuPetro}.

\subsection{Melting and differentiation}

The first melting process in chondritic material would be the eutectic melting in the Ni,Fe-FeS system. It is usually assumed that central temperatures in the ordinary chondrite (H, L, LL, EH, EL) parent bodies always stayed below the temperature required for the onset of melting. The reason for this assumption is the lack of ordinary chondrites with indications for melting or at least partial melting. Recently primitive achondrites have been detected, whose oxygen isotopic ratios plot within the corresponding fields of L and LL chondrites \citep{Vac17, Vac17b,Vac18}. In particular NWA 11042 was interpreted as being derived from the L chondrite parent body. Its texture is achondritic and its elemental composition resemble that of L chondrites, but it is depleted of siderophile elements compared to L chondrites and the Ni,Fe-metal content is very low. This is interpreted as that the material of NWA 11042 is derived from L chondrite material by melting and differentiation. \citet{Vac17} speculated that this fits within the frame of the thermal evolution of the parent body according to the ``onion-shell'' model, but requires higher central temperatures and a bigger parent body radius than existing models. On the other hand, \citet{WuH17} alternatively propose that NWA 11042 derives from a massive impact produced magma chamber on the L chondrite parent body.

Here we check, whether it is possible that the core region in our model of the parent body of L chondrites may be the source of primitive achondrites. This requires that the central temperature well exceeds the melting temperature of the Fe,Ni-FeS eutectic, but does not result in extended silicate melting, since only partial melting of the silicates is observed. The eutectic melting temperature of Ni,Fe-FeS at contact between Fe,Ni-metal and FeS grains was studied in \citet{Tom09} and \citet{Mar14}. According to them the addition of Ni to the Fe-FeS system lowers the eutectic melting temperature from 1\,261 K to 1\,223 K. In Figs.~\ref{FigTmax} and \ref{FigTmaxBig} we show the maximum temperature achieved during the thermal evolution of the body for each radius $r$ for models LM2 and  LM3, respectively. In our optimum model LM3 the melting temperature is, indeed, exceeded in a wide inner core region extending from the centre to about one half of the radius.\footnote{%
Melting is not consistently implemented in the simplified version of the model program used here, but since the central temperature in the model LT3 only barely exceeds the eutectic melting temperature, only small amounts of melt form which have no substantial influence on the model structure.}
Hence, the parent body of the L chondrites could have experienced incipient eutectic melting of FeS in its core region. This region would extend from the centre up to $r=82.0$~km which means that an inner core region of 13\% of the total volume would contain material that would be classified as a primitive achondrite. 

The absence of primitive achondrites related to L chondrites contradicts to such a model. Also the empirically determined upper metamorphic temperature of L6 chondrites of about $1\,173\pm30$ K given by \citet{Sla05} is in conflict with such a model, because the parent body of the L chondrites was most likely completely disrupted, in which case one should have meteoritic samples also from the deepest layers with temperature $\gtrsim1223$ K. The maximum metamorphic temperatures fit, however, reasonably well to the maximum temperature at the centre of 1186 K for our model LM2.


\begin{sidewaystable*}

\caption{Summary of model calculations for the thermal evolution of the L chondrite parent body.}

{\small
\begin{tabular}{lcccccccccc@{\hspace{.5cm}}l}
\hline
\hline
\noalign{\smallskip}
  & Mi81 & Be95 & \multicolumn{2}{c}{Be96} & BA02 & Bo07 & Sp11 & Ma14 & Bl18 & New & Units \\
  &      &      & por. & comp. &  &  &  &  &  &   \\
\noalign{\smallskip}
\hline
\noalign{\smallskip}
Method\tablefootmark{a} & Ana & Ana & \multicolumn{2}{c}{Ana} & Num & Num & Ana & Num & Ana & LM2 \\
Heat source & $^{26\!}$Al & $^{26\!}$Al & \multicolumn{2}{c}{$^{26\!}$Al} & $^{26\!}$Al  & $^{26\!}$Al, $^{60}$Fe  &  & $^{26\!}$Al  & $^{26\!}$Al, $^{60}$Fe & $^{26\!}$Al, \dots 
\\[.1cm]
  & \multicolumn{10}{c}{Input parameter} \\[.1cm]
Intrinsic density $\varrho_\mathrm{i}$ & 3\,610 & 3\,610 & 3\,650  & 3\,610 & --- &    & 3\,610 & --- & 3\,510  & 3\,580 & kg\,m$^{-3}$ \\
Bulk density $\varrho_\mathrm{b}$ & 3\,200 & 3\,610 & 3\,290 & 3\,520 & --- &    & 3\,200 & 3\,400 & 3\,350  & var & kg\,m$^{-3}$ \\
Porosity $\Phi_0$ & 11.3 & 0 & 10.3 & 2.5 & --- &   & 11.3 & 3.0 & 4.6 & var & \% \\ 
Heat conductivity core $K$ & 1.0 & 1.98 & 0.90 & 2.35 & var &    & 1.0 & 4.0 & 4.0 & var &  W\,m$^{-1}$K$^{-1}$ \\
--- mantle                 &     &      &      &      & var &    &     & 1.0 &     & var &  W\,m$^{-1}$K$^{-1}$ \\
Heat capacity $c_p$ & \emph{625} &  \emph{720} & \emph{730}  & \emph{760} & var &    & 625 & 872 & 930 & var &  J\,kg$^{-1}$K$^{-1}$ \\
$^{26}$Al/$^{27\!}$Al ratio\tablefootmark{b}   & $5\times10^{-5}$ & $5\times10^{-5}$ & $5\times10^{-5}$ & $5\times10^{-5}$ &  $5\times10^{-5}$  &    & $5\times10^{-5}$  & $5\times10^{-5}$ & $5\times10^{-5}$ & $5.25\times10^{-5}$ & \\
Heating rate\tablefootmark{c}  &  $1.82\times10^{-7}$ & $1.97\times10^{-7}$   & $1.97\times10^{-7}$  &  $1.97\times10^{-7}$  & $1.97\times10^{-7}$  &      & $1.82\times10^{-7}$  & $2.52\times10^{-7}$ & $3.37\times10^{-7}$ & $1.65\times10^{-7}$  &   W\,kg$^{-1}$  \\[.1cm]
  & \multicolumn{10}{c}{Initial and boundary condition}
\\[.1cm]

Initial Temperature $T_0$ & 180 &  180 & 300 & 300 &  200 &   & 300 &  180 & 250 & 200 & K \\
Boundary temperature $T_\mathrm{b}$ & 180 &  180 & 300 & 300 & 200 &    & 300 & 180 & 250 & 200 & K 
\\[.1cm]
  & \multicolumn{10}{c}{Parent body parameter}
\\[.1cm]
Radius $R$ & 85 &  60 & 54 & 90 & 100  &  100  & 100 & 100 & $>240$  & 115 & km \\
Birthtime $t_\mathrm{b}$ & 2.5 & $\sim 2$ & 1.6 & 2.0 & 2.3 & 2 -- 4  & 2.7 & 2.22  & 2.05 -- 2.25 & 1.89 &  Ma \\
Central temperature & 1\,150 &  1\,273 & 1\,273 & 1\,273 & $\sim$1\,230 &  $\sim1\,700$ & 1\,170 &  1\,273 & 1\,173 & 1\,186 & K \\
\noalign{\smallskip}
\hline
\noalign{\smallskip}
Reference & (1) & (2) & \multicolumn{2}{c}{(3)} & (4) & (5) & (6) & (7) & (8) & (9) \\
\noalign{\smallskip}
\hline
\end{tabular}
}
\tablebib{
(1)~\citet{Miy81}; (2) \citet{Ben95}; (3) \citet{Ben96}; (4) \citet{Ben02} ; (5) \citet{Bou07}; (6)\citet{Spr11}; (7) \citet{Mar14}; (8) \citet{Bla17}; (9) this work
}
\tablefoot{
\tablefoottext{a}{Ana: Analytic solution of the heat conduction equation, material properties constant. Num: Numerical solution of the heat conduction equation, material properties temperature dependent.}
\\
\tablefoottext{b}{At time of solar sytem formation}
\\
\tablefoottext{c}{By decay of $^{26}$Al at time of solar sytem (CAI) formation. In models Mi81, Be95, Be96, Sp11 the heating rate by $^{26\!}$Al is determined from the expression given in \citet{Her77}, in model Bl18 the heating rate is from \citet{Cas09}, and in the present paper it is calculated as described in \citet{Hen11}.}
}

\label{TabModls}
\end{sidewaystable*}

\section{Comparison with previous model calculations}

\label{SectComp}

\subsection{Method of modelling}

The number of published model calculations for the thermal evolution of the L chondrites parent body \citep{Miy81,Ben95,Ben96,Ben02,Bou07,Spr11,Mar14,Bla17} is small, probably because of the few available data on L chondrites suited for a comparison of models with laboratory data. The basic characteristics of the published models are summarized in Table~\ref{TabModls}, except for the model of \citet{Ben02} for which no details are published. Most of these models assume a short duration of the formation period of the  body compared to the characteristic heating time by $^{26\!}$Al such that the instantaneous formation hypothesis can be applied. Modern theories of planetesimal formation support this assumption \citep[e.g.][]{Joh15}. Only the model calculation of \citet{Bou07} includes growth of the L chondrite parent body over an extended period which results in a somewhat different evolution from that in the other models. In the following we consider these models in detail.

The model of \citet{Miy81} \citep[and the preliminary communication by][]{Miy80} was the first complete model calculation for L chondrites parent bodies, based on internal heating by the short-lived radio-nuclide $^{26\!}$Al (it also discusses the evolution of H chondrites). It follows essentially the method proposed by \citet{Min79} to calculate the thermal history of the H chondrite parent body. 

The model calculation is based on the analytic solution of the heat conduction equation (\ref{HeatCond}), as given in \citet{Car59}, for the case of spatially and temporal constant values of heat conductivity $K$, specific heat capacity $c_p$, density $\varrho$, and initial temperature $T_0$, and a spatially constant and exponentially decaying heat source. This analytic model forms also the basis of almost all subsequent models for the thermal evolution of the L chondrite parent body considered so far \citep{Ben95,Ben96, Spr11,Bla17}. The different models merely differ with respect to the particular choice of the constant input parameters. 

In the model calculations of \citet{Ben02}, \cite{Mar14}, and the present work the heat conduction equation is solved numerically. This relaxes all restrictions with respect to the variation of the coefficients and initial/boundary and allows to couple the evolution of temperature with other processes like melting \citep[in][]{Mar14} and sintering of the initially porous material (this work).

\subsection{Material properties}

The requirement that for the definitely temperature and porosity dependent parametres entering the heat conduction equation ($c_p$, $K$, $\varrho$)  spatially and temporal constant values have to be used in the analytic solution makes it necessary to chose some fixed value for these quantities which are thought to be representative for all zones of the body and during the complete heating and cooling period. This casts some doubts on the reliability of the models obtained by using the analytic solution because in the interior of the body the initially porous material will be compacted by sintering at temperatures $700\ \dots\ >900$ K (depending on the chondrule diameters), while in the cold outer layers the material remains loosely compacted. Hence the heat conductivity will be quite different in the core and in outer layers of the body. For this reason \citet{Ben96} calculated two different models with constant values of the material properties throughout the body as limit cases, one for a strongly compacted material (tagged by `por.' in Table \ref{TabModls}) and another one for a porous material (tagged by `comp.' in Table \ref{TabModls}), assuming that a solution accounting for the variability of material properties would be found somewhere in between. For H chondrites two-zone models with a conductive core and an insulating mantle to handle this problem have been studied by \citet{Akr98} and \citet{Har10}. For L chondrites such type of study  has been conducted only by \citet{Ben02}, but no details are given for that model.

The density is chosen in all models according to observations on L chondrites.  \citet{Yom83} give the average chemical composition of L chondrites, which is also adopted in our model calculation. From this they calculate an intrinsic density, $\varrho_\mathrm{i}$, of 3\,610 kg\,m$^{-3}$. This value is used in \citet{Miy81}, \citet{Ben95}, the compacted model of \citet{Ben96}, and in our model. For the bulk density, $\varrho_\mathrm{b}$, the value of 3\,200 kg\,m$^{-3}$ used in \citet{Miy81} and \citet{Spr11} is the average of the H chondrites from Antartica listed in \citet{Yom83}. This corresponds to an average porosity of about 11\%, a value that deviates from values given in \citet{Bri03} for a much bigger sample, but since the solution depends only weakly on density its assumed value is not so important for the results. The bulk densities and intrinsic densities in \citet{Ben96} for their two models correspond to the data of the two meteorites Y750977 (porous material) and Arapahoe (compacted material) considered as representative for loosely and strongly compacted material. The bulk density used in \citet{Mar14} is taken from the compilation in \citet{Wil03}, no value for $\varrho_\mathrm{i}$ or porosity is given. In \citet{Bla17} values for the average bulk density and the intrinsic density are taken from \citet{Bri03}. The density used in \citet{Ben02} is not specified.

\begin{figure}

\includegraphics[width=\hsize]{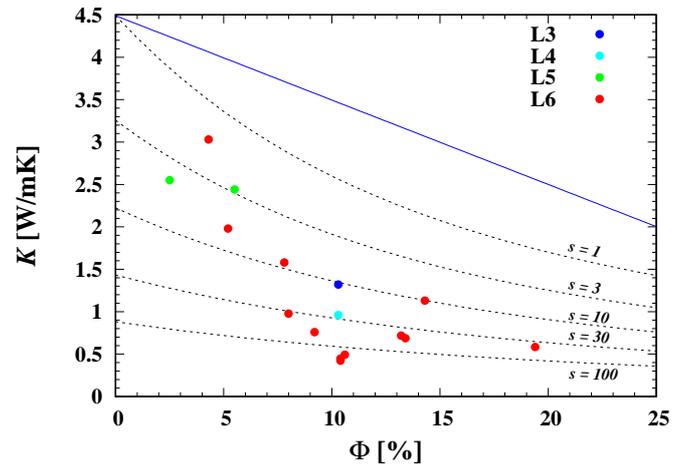}

\caption{Measured heat conductivities and porosities of L chondrites at 300 K \citep{Yom83,Ope10,Ope12} for which both quantities have been measured. The dashed lines correspond to the heat conductivity according to the conduction model (Sect.~\ref{SectHeatCond}) for a porous material with abundant micro-cracks. The indicated values correspond to the reduction factor, $s$, for phonon scattering length. The blue line correspond to the heat conductivity of a sand-stone like porous material according to Eq.~(\ref{HeatCondPor}). }

\label{FigKchonL}
\end{figure}

A significant variation with porosity is observed for the heat conductivity $K$, which has a strong impact on the characteristic cooling timescale. Measured values for L chondrites show a wide and unsystematic variation with porosity (between $\sim0.4$ and $\sim3$ W\,m$^{-1}$K$^{-1}$ at 300 K, see Fig.~\ref{FigKchonL}, which is about a factor of about eight between lowest and highest value) and quite different types of temperature variation \citep{Yom83, Ope10, Ope12}. The origin of these differences could be in \citet{Hen16} and \citet{Gai18}. The published model calculations for thermal evolution assume quite different values for the heat conductivity $K$, ranging between the lower end of observed values for L chondrites in the models Mi81, Be96(por), and Sp11, up to the value of the fully compacted material in Bl18 \citep[4.17 W\,m$^{-1}$K$^{-1}$, see ][]{Hen16}, and a more intermediate value in Be95 and Be96(comp). Using the low values for the heat conductivity means that more stress is laid on the heat transport in the surface-near layers which is responsible for the global evolution of the heat content of the body, while high values better account for the temperature structure in the interior of the body where the material is compacted by sintering. 
 
An equation for the global heat content of the body is obtained by integrating the heat conduction equation (\ref{HeatCond}) over the whole volume of the body
\begin{equation}
{\mathrm{d}\,U\over\mathrm{d}\,t}=-4\pi R^2K\left.{\partial\,T\over\partial r}\right\vert_{r=R}+M\cdot h\,,
\end{equation}
where $M$ is the total mass of the body, $U$ its total heat content, and $h$ the specific heating rate per unit mass. This shows that the long-term thermal evolution is crucially determined by the heat conductivity of the surface layers and that there is some justification to use a representative value for this in the model calculations. A value corresponding to the conductivity of the compacted material as used in model Bl18 in any case overestimates the rate of heat loss. The lower values chosen in the models Mi81, Be95, Be96, and Sp11 seem more appropriate, but the precise reasons which lead the authors to their preferred choice are not communicated in the papers. A good deal of the significant discrepancies between the radii found for the parent body of the L chondrites in the different models rests on the different choices for $K$. In particular the big radius found in model Bl18 is required to compensate for the high rate of heat lossy. 

Much less disagreement exists between the different choices for a representative value for the heat capacity in models Mi81, Sp11, Be95, Be96, and Bl18, see Table~\ref{TabModls}. Values given in italics are not explicitly stated in the papers but calculated from the values given for heat diffusivity, $\kappa=K/\varrho_\mathrm{b}c_p$, and heat conductivity $K$. Chosen values are in the middle range between a lower value of $\sim200$ J\,kg$^{-1}$K$^{-1}$ at a temperature of 200 K at the surface and the highest value of $\sim 1200$ J\,kg$^{-1}$K$^{-1}$ at a peak temperature of $\sim1\,200$ K achieved at the centre. Also in the case of heat capacities the reasons which lead the different authors to their preferred choice are not communicated in the papers.  A variation of the cooling timescale resulting from the temperature variation of the heat capacity is moderate and, thus, much less influential for the model than the variation resulting from variations in the heat conductivity. 

\begin{figure}

\includegraphics[width=\hsize]{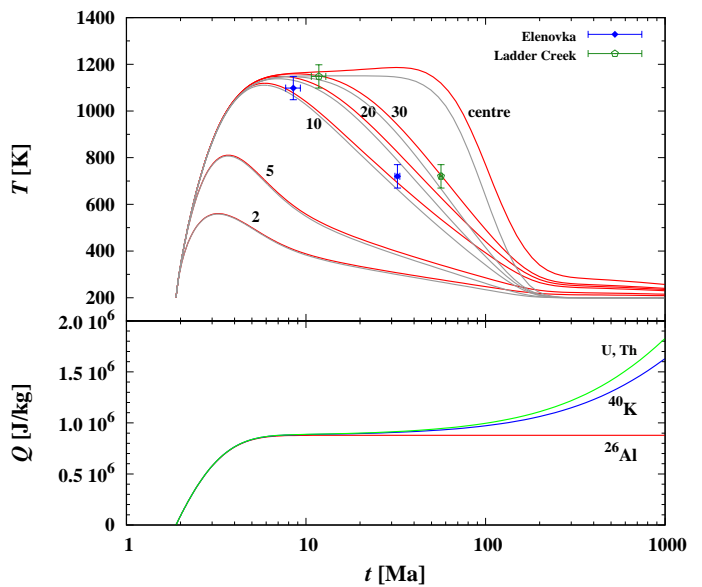}

\caption{Upper panel: Temperature evolution of the parent body model at a set of selected depths (km) considering the contributions of the full set of heat sources (red lines) for our optimized model, and a model with the same formation time and radius but with the sole contribution of $^{26}$Al to heating (grey lines). Also shown for comparison are Hf-W and Pb-Pb ages and the corresponding closure temperatures for the two indicated meteorites. Lower panel: Cumulated heat released, $Q$, up to instant $t$ for the different contributing radioactive species.
}

\label{FigHeat26Al40K}
\end{figure}

Most of the models considered assume that the heating of the body is due to the decay of radioactive nuclei and that the heating is dominated by the decay of $^{26\!}$Al such that it suffices to consider only this contribution. A possible contribution of $^{60}$Fe to the heat budget of the body was considered in \citet{Bou07} and \citet{Mar14}. Its importance depends on the assumed initial abundance of $^{60}$Fe, which is disputed, and on the half-life of $^{60}$Fe which was corrected in \citet{Rug09}. Older determinations of the $^{60}$Fe/$^{56}$Fe ratio found a value of $9.2\times10^{-7}$ \citep[e.g.][]{Mos05} which was used in the model of \citet[ which also used an outdated value for the half-life]{Bou07}, while more recent determinations gave a much lower vale of $1.15\times10^{-8}$ \citep{Tan15} which is used in \citet{Mar14} and in our calculation. With the low value of the $^{60}$Fe/$^{56}$Fe ratio the contribution of  $^{60}$Fe to the heating is not important. This is in accord with our finding for the H chondrite parent body that if the $^{60}$Fe abundance is considered as a free parameter and determined from a best fit of models to meteoritic data then it is found that the observed thermal evolution is best reproduced without a contribution of $^{60}$Fe to heating \citep{Hen13}.  

There is, however, some contribution to heating of the body by long-lived radioactives, in particular by $^{40}$K. For asteroid-sized bodies the main effect of these heat soures is a certain delay of the onset of cooling in the central region of the body. This manifests in the slight increase of the central temperature during the first 100 Ma recognizable in Fig.~\ref{FigFitMet}. These heat sources are negected in all previous model calculations because they are not important for bodies of the 100 km size-class, but are included in our model to allow also the modelling of bigger-than-usual bodies for which they may become important.   
 
\subsection{Contribution of different heat sources}

Figure \ref{FigHeat26Al40K} demonstrates the effect of including other heat sources than $^{26\!}$Al by comparing the temperature evolution at some selected depths below the surface for our preferred model LM2 with that in a model where only $^{26\!}$Al is considered. During the cooling phase of the body, there are moderate, but significant differences of the instant at which the temperature at a depth drops below a given temperature, e.g., the closure temperature of a thermochronological system. The difference between the two models is essentially due to the contribution of $^{40}$K; the contribution of U and Th becomes only important at times where temperatures dropped already well below the closure temperatures of the thermochronological systems of interest, and $^{60}$Fe is never important at the presently favoured low abundance in the solar nebula. For accurate thermochronological data (as those we used) the accuracy with which closure times for a specific radiometric clock are determined are much better than the difference between the instances where the cooling curve for a meteorite passes through the corresponding closure temperature for different parent body models that either include heating by $^{40}$K or not. If detailed models for thermal evolution of asteroids are compared to such accurate data, the contribution of $^{40}$K has to be considered. If, on the other hand, more qualitative models based on a solution of the heat conduction equation with constant coeffcients are considered, it suffices to consider only $^{26\!}$Al as heat source. 

\begin{table}

\caption{Model results for the percentage of volume in the parent body of L chondrites occupied by material that suffered the range of metamorphic temperatures corresponding to petrologic types three to six for different models, observed percentages for H meteorites (falls and falls + finds), and maximum metamorphic temperatures of the petrologic types used by the different authors for calculating volume fractions.
}

\begin{tabular}{lc@{\hspace{.3cm}}rrrrl}
\hline
\hline
\noalign{\medskip}
Model\tablefootmark{c}  & & L3 & L4 & L5 & L6 \\
\noalign{\smallskip}
\hline
\noalign{\smallskip}
Mi81     & & 8.6  & 14.2 & 28.3 & 49.0 & Vol\% \\
         & & 500  &  825 & 925  &    & K  \\[.1cm]    
\multicolumn{2}{l}{Be96comp\tablefootmark{b}}  & 22.5 & 7.0 & 3.8 & 66.7 & Vol\% \\
         & & 873  & 973  & 1023 & 1298 & K \\[.1cm]
Ma14\tablefootmark{c}   
         & & 14\phantom{.0}  & \multicolumn{2}{c}{5} & 81\phantom{.0} & Vol\%  \\
         & & 948  & \multicolumn{2}{c}{1138} &  1273 & K \\[.1cm]
LM2      & & 15.6 & 6.7 & 15.8  & 61.8 & Vol\% \\
         & & 873  & 973 & 1100  &    & K  \\[.1cm]
Observed\tablefootmark{d} & (falls) & 3.9 & 7.2 & 21.9 & 67.2 & Vol\% \\
         & (all) & 5.8  & 20.1 & 31.1 & 42.8 & Vol\% \\[.3cm]
\hline
\noalign{\medskip}
      & & H3 & H4 & H5 & H6 \\
\noalign{\smallskip}
\hline
\noalign{\smallskip}
Mi81     & & 8.6  & 17.1 & 47.2 & 27.1 & Vol\% \\
         & & 500  &  825 & 925  &      & K     \\[.1cm]    
\multicolumn{2}{l}{Be96comp\tablefootmark{b}} & 19.1 & 5.9  & 3.3  & 71.7 & Vol\% \\
         & & 873  & 973  & 1023 & 1273 & K     \\[.1cm]
Ma14\tablefootmark{c}   
         & & 16\phantom{.0}  & \multicolumn{2}{c}{7} & 77\phantom{.0} & Vol\%   \\
         & & 948  & \multicolumn{2}{c}{1138} &  1273 & K \\[.1cm]
HM       & & 14.3 & 3.5 & 4.1 & 71.8 & Vol\% \\
         & & 873  & 973 & 1075  &    & K  \\[.1cm]
Observed & (falls) & 5.2 & 18.6 & 50.5 & 25.7 & Vol\%  \\
    &(all)         & 5.5 & 27.6 & 42.3 & 24.6 & Vol\%  \\
\noalign{\smallskip}
\hline
\end{tabular}
\tablefoot{\small
\tablefoottext{a}{Key to models see Table \ref{TabModls}. Model HM is for the H chondrite parent body determined in the same way as described in this paper, but using data for H chondritic material and meteoritic data as given in \citet{Hen13}.}
\tablefoottext{b}{Values correspond to the model of the compacted body which is closest to our preferred model.}
\tablefoottext{c}{Estimated from their Fig.~9. Petrologic type four and five are not discriminated.}
\tablefoottext{d}{Data from the Meteoritical Bulletin Databasis {\tt https://www.lpi.usra.edu/meteor/metbull.php}.}
}

\label{TabVolFracPet}

\end{table}

\subsection{Abundance of petrologic types}
\label{SectAbuPetro}

It is generally assumed that the metamorphism of the chondritic material from the highly pristine material seen in L3 meteorites to the highly equilibrated material seen in L6 meteorites is essentially determined by the peak temperature achieved by the material during the thermal evolution of the parent body. For given maximum metamorphic temperature of a particular petrologic type and a temperature evolution model of the parent body one readily derives the depth of the borders between the zones where the structure of the material corresponds to a given petrologic type and the next higher one, and then the volume percentages of the parent parent body occupied by the different petrologic types.   

\citet{Miy81} derived in this way the volume fractions of L3 to L6 material in the parent body of the L chondrites and compared this to the observed fall statistics of L chondrites. 
They used a value of 500~K for the maximum metamorphic temperature of L3 chondrites, a maximum temperature of 800--850~K for type L4, and a maximum temperature of 1\,150 K for type L6. The border between type 5 and type 6 was somewhat arbitrarily set to the point of steepest descent of the $T_\mathrm{max}$ curve.

The calculated volume fraction of L3 was used by \citet{Miy81} as an additonal information to fix the boundary temperature $T_\mathrm{b}$ of the model, because obviously the volume fraction of L3 material in the body strongly depends on the surface temperature. The observed volume fraction of 5--10\% L3 chondrites from all antarctic L chondrites motivated the choice of a surface temperature of 200~K for their models.  

\citet{Ben96} used values of maximum metamorphic temperatures of 873 K, 973 K, and 1023 K for L3, L4, and L5 chondrites, respectively, from \citet{McS88}, and of 1273 K for L6 chondrites from \citet{Har93} to calculate volume fractions. \citet{Mar14} used the values for maximum metamorphic temperature as proposed by \citet{Har10}: 948 K for type 3 and 1273 K for type 6. They do not discriminate between types four and five; the boundary between type 5 and type 6 is assumed at 1138 K. The data we adopt for the maximum metamorphic temperatures to calculate volume fractions for our mode LM2 are shown in Table \ref{TabMaxTempPetro}. The resulting volume fractions for this and the previous models are given in Table \ref{TabVolFracPet} where they are compared to the fractional abundance of L3 to L6 chondrites as derived from the numbers of known meteorites of type L3 to L6. For comparison we also show the corresponding numbers for the case of the H chondrites, the only other meteorite class for which such information is presently available. 

In all cases, both for L and H chondrites, no good agreement was found between calculated and observed volume fractions. This is no surprise because the observed relative proportions of the different petrologic types depend on the eventualities of the collision history of large to small bodies and of collision fragments in the planetary system during the last 100 million years or so and on diverse biases for recovering of meteorites of different types. It is presently not possible to reconstruct the volume percentages of petrologic types three to six in the parent body from that observed for meteorites.


\section{Concluding remarks}
\label{SectConclu}

We attempted to reconstruct the parent body of the L chondrites from empirical data on the cooling history of meteorites. We collected from literature the available data on closure times for different thermochronological systems and for cooling rates. The number of useful data found for L chondrites is far less than for the much better studied H chondrites, despite the fact that L chondrites are the most frequent ordinary chondrites found. This certainly results from the fact that the parent body of the L chondrites catastrophically disrupted 470 Ma ago by a collision that is responsible for the high concentration of fossil meteorites in mid-Ordovician marine limestone in southern Sweden \citep{Hec04,Hec08} and that only few L chondrites are found for which the radioactive decay clocks are not reset by this event. Totally, for five meteorites high-quality thermochronological data for two thermochronological systems, the Hf-W and U-Pb-Pb systems, are available. This is just sufficient observational material to pin down the two most important parameters of the parent body, its radius and its formation time, by fitting models for the internal constitution and evolution of small bodies from the Asteroid belt to the meteoritic record. Some additional data material on the cooling history of L chondrites is available that can be compared with parent body models, but is of insufficient accuracy that it can be used to narrow down the parent body properties.

Based on the assumption of a spherically symmetric body that formed instantaneously and is heated by decay of long and short lived radioactives, in particular by $^{26\!}$Al, we solved the heat conduction equation coupled with equations for the sintering of the initial granular chondrule-matrix mixture and for the porosity and temperature dependence of the heat conductivity. Additionally we allowed for a surface layer of reduced heat conductivity due to impact-generated cracks. By varying the radius and formation time (after CAI formation) we determined the parameter combination which fitted the set of high-quality thermochronological data as good as possible. 

It turned out that two different types of models are compatible with the available data. One solution is a body with radius of 115 km and formation time of 1.89 Ma after CAI formation, another solution is a body with 160 km radius and 1.835 Ma formation time. The basic difference between the alternative models is that for the bigger model the core region of the body shows incipient eutectic melting of Ni,Fe-FeS, while for the smaller model the central temperature remains well below the melting temperature. The lack of primitive achondrites that are related on compositional grounds to L chondrites presently favours the smaller of the two models.


\begin{acknowledgements}
We acknowledge J. L. Hellmann, T. S. Kruijer, J. A. Van Orman, K. Metzler, and T. Kleine for making available to us their data on Hf-W ages and cooling of L chondrites prior to publication. This work was performed as part of a project of `Schwerpunktprogramm 1385', supported by the `Deutsche Forschungs\-gemeinschaft (DFG)'.  MT acknowledges support by the Klaus Tschira Stiftung gGmbH. This research has made use of NASA's Astrophysics Data System.
\end{acknowledgements}


\begin{appendix}

\section{Properties of chondritic material}
\label{AppMatProp}

\subsection{Porosity}
\label{SectIniPor}

The composition and structure of the material of the parent body is inferred from the properties of the meteorites. The material of the L chondrites is a mixture of millimetre-sized chondrules embedded in a matrix of micrometre-sized dust grains. This material is metamorphosed to varying degrees by action of high temperatures and  pressures due to internal heating of the body and the action of impacts on surface-near material. Its properties form a sequence between a weakly consolidated and highly porous material composed of clearly discernable chondrules and matrix (petrologic type 3) and a higly compacted and re-crystallised material with hardly discernable chondrules (petrologic type~6), with all possible intermediate states between the two extreme cases.

It is assumed that the most porous material found in L3 chondrites represents the composition and structure of the material found in surface-near layers of the parent body and essentially represents the initial granular state of the parent body material. The consolidated and crystallised material found in L5 and L6 chondrites is assumed to be the product of sintering and crystal growth at high temperature and moderate pressure during transient heating in deeper layers of the parent body, resulting from the same kind of material as found in L3 chondrites. 

The material of chondrites originating from surface-near layers of the parent body, however, does not simply represent the pristine structure of the parent body material. During the $\sim4.6$ Ga period between formation of the body and the excavation of a meteorite from the surface of the body or of a major fragment of it, material from the first few km beneath the surface has been subject to many shock waves emanated by impacts of small to large bodies from the Asteroid belt onto the parent body. If the peak pressure in the shock wave exceeds about 5 GPa \citep{Sto91} existing pore space in the material is reduced or closed (by rolling, gliding, crushing, transient melting), while at the same time numerous micro-cracks are freshly generated in the mineral crystals. 

\begin{table}

\caption{Basic parameters used for the model calculation}

\begin{tabular}{llll}
\hline
\hline
\noalign{\smallskip}
Quantity & Symbol & Value & Unit\\
\noalign{\smallskip}
\hline
\noalign{\smallskip}
Av. chondrule radius           &              & 0.25                 & mm            \\
Av. matrix grain radius        &              & 1                    & $\mu$m        \\
\\[.1cm]
%
Intrinsic mass density         & $\varrho_\mathrm{i}$  & 3.58        & g\,cm$^{-3}$  \\
Mass fraction Fe               & $X_{\rm Fe}$ & $2.21\times10^{-1}$  &               \\
Mass fraction Al               & $X_{\rm Al}$ & $1.23\times10^{-2}$  &               \\
Mass fraction K                & $X_{\rm K}$  & $9.96\times10^{-4}$  &               
%
\\[.1cm]
Heat conductivity\tablefootmark{a}        & $K_b$        & 4.29                 & W\,m$^{-1}$\,K$^{-1}$ \\
Specific heat\tablefootmark{a}            & $c_p$        & 714                  & J\,kg$^{-1}$\,K$^{-1}$ 
\\[.1cm]
$^{26\!}$Al/$^{27\!}$Al ratio\tablefootmark{b}  &              & $5.25\times10^{-5}$  &               \\
$^{60}$Fe/$^{56}$Fe ratio\tablefootmark{b}    &              & $1.15\times10^{-8}$  &               \\
$^{40}$K/$^{39}$K ratio\tablefootmark{b}        &              & $1.58\times10^{-3}$  &                \\
Heating rate $^{26\!}$Al       & h            & $1.65\times10^{-7}$  & W\,kg$^{-1}$  \\
\quad Half live                & $\tau_{1/2}$ & 0.72                 & Ma            \\
\quad Energy                   & $E$          & 3.19                 & MeV           \\
Heating rate $^{60}$Fe         & h            & $9.10\times10^{-11}$ & W\,kg$^{-1}$  \\
\quad Half live                & $\tau_{1/2}$ & 2.62                 & Ma            \\
\quad Energy                   & $E$          & 2.89                 & MeV           \\
Heating rate $^{40}$K          & h            & $3.11\times10^{-11}$ & W\,kg$^{-1}$  \\
\quad Half live                & $\tau_{1/2}$ & 1248                 & Ma            \\
\quad Energy                   & $E$          & 0.693                & MeV           \\
\noalign{\smallskip}
\hline
\end{tabular}
\tablefoot{
\\
\tablefoottext{a}{At 300 K. $K_\mathrm{b}$ is for the completely compacted material.}
\\
\tablefoottext{b}{Data for $^{26\!}$Al from \citet{Kit13}, for $^{60}$Fe from \citet{Tan15}, and for $^{40}$K from \citet{Lod09}.}
}

\label{TabModDat}
\end{table}

In order to determine the initial porosity of the parent body material, we have to look for the highest porosities observed in meteorites which show the least indications of shock metamorphosis \citep[shock stage S1, cf.][]{Sto91} because these also show the highest porosities \citep{Con08}. These high porosities are not necessarily found, however, amongst the chondrites of the lowest petrologic type, because they are just those most affected by impact compaction. Intermediate types (L4,L5) are more promising for this, while type L6 meteorites are already heavily compacted by sintering.  The investigations of porosity of meteorites by \citet{Con98} and \citet{Con08} found a broad distribution of porosities extending from nearly zero values up to $\sim0.2\dots0.25$ with some individuals showing porosities up to 0.3. On average, the porosities are highest for the lowest shock stage, as expected. We interpret the findings in \citet{Con98} as showing the important role played by impacts and sintering on a material with initially porosity corresponding to the highest observed values. We take a porosity of $\Phi=0.25$ as the initial value of the average porosity of the pristine material of the L chondrite parent body.

This value of $\Phi$ fits well to the porosity of $\Phi_0=0.248$ expected for a binary granular mixture consisting of two components with significantly distinct diameters of the granular units of the two components \citep[cf. ][]{Gai15}. For the L chondrites the chondrules have typical diameter of 0.5 mm with a rather narrow distribution around the mean \citep{Fri15} while the matrix particles have diameters of the order of a few micro-metres (we chose a value 2 $\mu$m). These values are used as initial values for the calculation of the porosity. 
  
\subsection{Composition}

With respect to the average composition of the main components of the mineral mixture of L chondrites we assume modal mineral abundances as given in \citet{Yom83} which are almost the same as given in the more recent compilation in \citet{Sch07}. This composition is used to calculate the heat capacity and heat conductivity of the material. 

For calculating the heat production by decay of radioactives we need to know the average mass fraction of the corresponding elements. For the elements Al, K, and Fe we use data for the composition of L chondrites from Table 1 in \citet{Sch07} from which we calculate the mass fractions $X_i$ of the elements in L chondrites given in Table \ref{TabModDat}. These are used by us to calculate the heat production rate.

The mass fractions of elements calculated for the composition in \citet{Yom83} are slightly different from that given in \citet{Was88} and from the values calculated from \citet{Sch07}, but usually within 20\% which is less than the uncertainty of the averages. We prefer the values derived from \citet{Sch07} because they are based on a broader data set.

\subsection{Heat production}

For the specific heat production rate, $h$, one has to consider the contributions of short- and long-lived radioactives, in particular that of $^{26}$Al, $^{60}$Fe, $^{40}$K which determine the temperature evolution during the first 200 Ma of evolution which are of main interest for us. For the evolution on longer terms one also has to consider Th and U. The heating rate $h$ in the heat conduction equation has the form
\begin{align}
h&=\sum_ih_i\,{\rm e}^{-\lambda_i(t-t_0)}\,.
\end{align}
The time $t_0$ is the time of solar system formation which is identified with the time of CAI formation. The quantities $\lambda_i$ are the decay constants of the nuclei and 
\begin{align}
h_i&={X_i\over mi}\,f_i\,\lambda_iE_i
\end{align}
the specific heating rates as calculated from the composition of the chondritic material and the initial abundances of the radioactive nuclei at time $t_0$. The decay constants are $\lambda_i=\ln 2/\tau_{1/2,i}$, where $\tau_{1/2,i}$ is the half live of the isotopes of interest, $X_i$ is the mass fraction of the corresponding element and $m_i$ its atomic mass, $f_i$ the fraction of the isotope of the isotope at the time of solar system formation, and $E_i$ the energy released (without neutrino energies) by the decay into the final stable state. The sum runs over the different radioactive nuclear species which are important for heating. The heating rates $h_i$ do not depend on time and temperature. They also do not depend on $r$ as long as there occurs no differentiation. The values of the coefficients used in the calculation are shown in Table~\ref{TabModDat}.   

\begin{figure}
\includegraphics[width=\hsize]{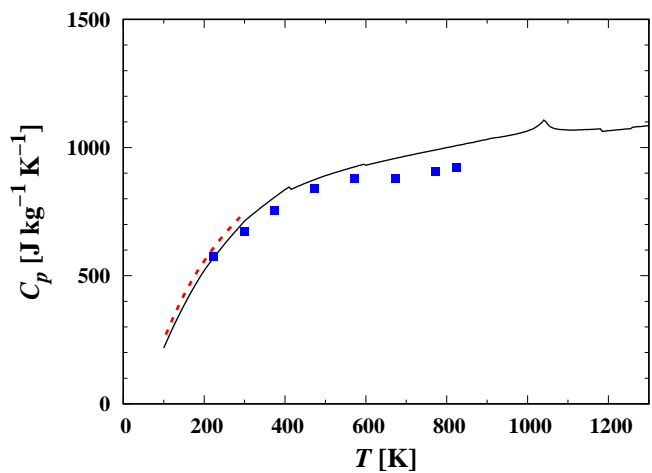}
\caption{Specific heat of L chondrite material, as calculated from average composition (black line), average values of experimental data (red dashed line) according to \citet{Mak16}, and data measured for an L6 chondrite \citep{Wach13} (blue squares). The discontinuities and the cusp at certain temperatures are due to phase transitions in some components of the mixture.}

\label{FigCv}
\end{figure}

The heat production rate may also include other heat sources, e.g., the latent heat during  melting and crystallisation, but apparently this is not of interest for the L chondrite parent body and we neglect them.

\subsection{Heat capacity and conductivity}
\label{SectHeatCond}

For the kind of investigations which we have in mind it is important to base a calculation of the temperature evolution on heat capacities, $c_p$, and heat conductivities, $K$, which take account of their temperature dependencies. The temperature structures of models calculated with constant values of $c_p$ and $K$ deviate significantly from models with realistic temperature variation and porosity dependence, as was already found by \citet{Yom84} and  demonstrated in detail by \citet{Gho99} and \citet{Akr98}, and was also found to in our previous model calculations. Even in a simplified model calculation one has to observe this.

\subsubsection{Heat capacity}

The specific heat capacity $c_p(T)$ of the chondritic mixture of minerals and metal is calculated from the specific heat capacity $c_{p,i}(T)$ of the components $i$ and their mass fractions $X_i$ in the mixture found in L chondrites as
\begin{equation}
c_p(T)=\sum_iX_i\,c_{p,i}(T)\,.
\label{EqCv}
\end{equation} 
The heat capacity is calculated for the standard pressure of one bar since for the pressure range of concern for bodies not bigger than a few hundred kilometre radius $c_p$ is practically pressure independent. Further details are given in \citet{Hen11}. The result is shown in Fig.~\ref{FigCv} which shows for comparison also the result of a laboratory measurement of $c_p$ for $T\ge300$ K of the L6 chondrite So{\l}tmany \citet{Wach13} and for some L chondrites for $T\le300$ K by \citet{Mak16}. We see reasonable agreement between $c_p$ calculated according to Eq.~(\ref{EqCv}) and measured heat capacities.

In the temperature range $T>300$ K some discontinuities and a cusp are found in the temperature variation of $c_p$. These are due to phase transitions (mainly in the FeS component) and the Curie point of iron. The heat capacity shown in Fig.~\ref{FigCv} does not include a contribution of molten components, because we assume that the temperature at the centre of the L chondrite parent body stays below the Fe,Ni-FeS eutectic melt temperature of 1\,223~K at the Ni content of L chondrites \citep{Tom09}. 

In the model calculation first a table for $c_p(T)$ is generated and during the model calculation the value of $c_p(T)$ at the required temperature is determined by interpolation from the table.

\subsubsection{Heat conductivity}
\label{SectHeatCond1}

The temperature dependence of the heat conductivity of chondrites was studied in \citet{Gai18}. We found that the experimental findings for the heat conductivity of chondrites \citep{Yom83,Ope10,Ope12} can be explained rather well by the heat conductivities of the individual components of the mixture if one appropriately considers the influence of isolated pores and in particular the network of abundant micro-cracks in the silicate mineral components. The model accounts for the observation that the observed heat conductivity of meteorites is far below that of the bulk material and it explains the large differences in the observed temperature variation of heat conductivity in meteorites of otherwise similar properties. 

\begin{figure}

\includegraphics[width=\hsize]{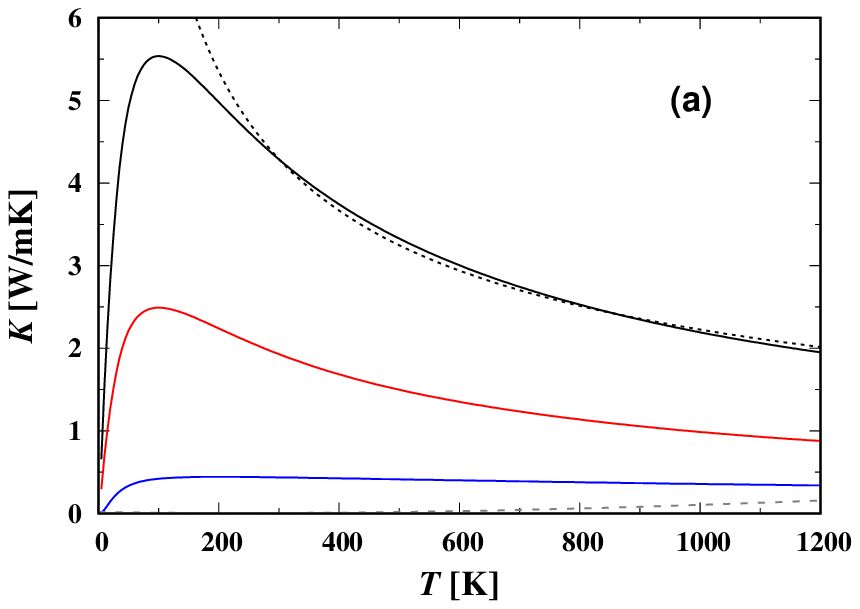}

\includegraphics[width=\hsize]{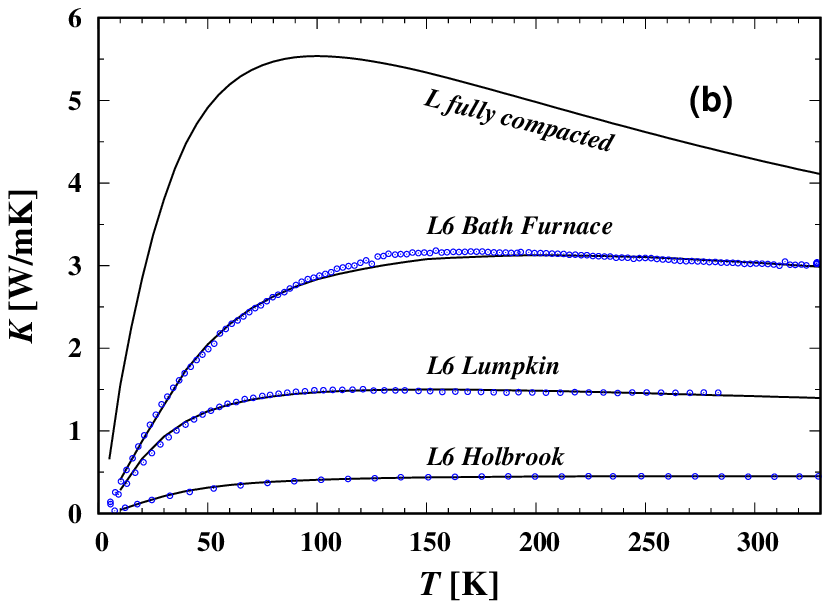}

\caption{(a) ) Heat conductivity by phonon-transport of the L chondrite bulk material, as calculated from its composition (black line), conductivity for a porous material with a 25\% porosity (red line), and conductivity for material with increased phonon scattering by micro-cracks (blue line) with phonon scattering length reduced by a factor $s=30$. The dashed line shows the contribution of radiative transfer to heat conductivity.
(b) Comparison of the theoretical model \citep{Gai18} for L chondrites (black lines) to measured heat conductivities of three L6 chondrites from \citet{Ope10,Ope12} (blue circles) by fitting the model parameters porosity (due to isolated pores) and scattering enhancement factor (accounting for micro-cracks).}

\label{FigKheat}
\end{figure}

Figure \ref{FigKheat}a shows the calculated effective heat conduction coefficient of chondritic material according to the model of \citet{Gai18} with composition corresponding to L chondrites for the compact material (solid black line), for a material with 25\% porosity by macro-pores (solid red line), and for a material where additionally the scattering of phonons in the silicate minerals is enhanced by numerous micro-cracks (solid blue line). 

Here we used a modified fit to the heat conductivity data of enstatite. The corresponding coefficients for the Callaway model \citep[see][ for their meaning]{Gai18} of phonon heat conductivity are given in Table \ref{TabEnst}. This reduces the deviation between the model and the laboratory data that remained in \citet{Gai18} for the case of the two enstatite chondrites from \citet{Ope10,Ope12} . With this modification it is possible to obtain a nearly perfect fit of the heat conductivities of L chondrites. Figure \ref{FigKheat}b shows measured heat conductivities in the temperature range from 10 to 300 K for three L chondrites from \citet[][ data made kindly available by G. Consolmagno]{Ope10,Ope12} and the temperature variation according to the model of \citet{Gai18} if the two basic parameters of the model, the porosity due to isolated pores, denoted by $\Phi$, and that due to micro-cracks, are determined by a fit to the empirical data. The close fit between model and measured data suggests that the theoretical model is suited to describe the temperature variation of the heat conductivity over the temperature range between about 20 K and 1200 K which is relevant for the modelling of the thermal structure of ordinary chondrite parent bodies, if the porosity, $\Phi$, due to macro-pores, the enhancement factor, $s$, for phonon scattering \citep[defined in ][]{Gai18}, and the aspect ratio, $\alpha$, of the pores are chosen appropriately.

\begin{table}
\caption{Coefficients for the Callaway-model of phonon heat conductivity for enstatite}
\begin{tabular}{lllll}
\hline
\hline
\noalign{\smallskip}
          & $p_1$ & $p_2$ & $K_0$ & $\Theta_\mathrm{a}$ \\
\noalign{\smallskip}
\hline
\noalign{\smallskip}
Enstatite & 1.0883 & $2.8128\times10^{-2}$ & $2.6293\times10^{1}$ & 160 \\
\noalign{\smallskip}
\hline
\end{tabular}

\label{TabEnst}
\end{table}

The most important factor determining the strong scatter of observed heat conductivities was found to be the reduction of phonon-scattering length by micro-cracks. Such micro-cracks are abundant in the present-day meteorites because of the many Ga lasting history of weak to strong impacts on the parent body. It is unlikely that they were already abundant during the early evolution period of asteroids, because during planetesimal formation mass is either acquired by low-velocity ($<1$ km\,s$^{-1}$ relative velocity) planetesimal-planetesimal collisions, if this should be the dominating formation route, or by low velocity acquisition of pebbles, if this would be the dominating formation route. In our model calculation we assume that initially only the macro-prosity due to the formation of the parent body from a granular material of chondrules and matrix is responsible for a reduction of the heat conductivity compared to that of the completely compacted material. In \citet{Hen16} it was found that the heat conductivity in porous material with macro-pores and a porosity not bigger than $\sim0.3$ can be approximated by the following expression
\begin{equation}
K(T,\Phi)=K_\mathrm{b}(T)\left(1-2.216\,\Phi\right)\,,
\label{HeatCondPor}
\end{equation}  
where $K_\mathrm{b}$ is the heat conductivity of the bulk material (Fig.~\ref{FigKheat}a, the black line). This relation is used to calculate the heat conductivity in Eq.~(\ref{HeatCond}). For the initial porosity $\Phi_0=0.25$ this means for instance that the heat conductivity is reduced to 45\% (the red curve in Fig.~\ref{FigKheat}a) of its value for the compacted material (the black curve in Fig.~\ref{FigKheat}a). The gradual vanishing of this macro-porosity at elevated temperature due to sintering is calculated as part of the model calculation.

In the model calculation first a table for $K_\mathrm{b}(T)$ is generated and during the model calculation the value of $K_\mathrm{b}$ at the required temperature is determined by interpolation from the table.

\subsubsection{Regolith layer}

After formation of Jupiter, relative velocities between planetesimals are pumped up to several km\,s$^{-1}$ \citep[e.g.][]{Dav13} and such  collisions would damage the mineral structure and generate abundant micro-cracks at the impact locations. This requires peak pressures $\gtrsim 5$ GPa at impact \citep{Sto91}. By this a surface layer of some thickness $d$ builds up where the material structure is is strongly modified by impacts. In this layer, the heat conductivity probably approaches the low values observed for present day asteroids (the blue curve in Fig.~\ref{FigKheat}a). 

But even then it would last considerable time until most of the surface of the body is affected by this and the temperature evolution of the body is modified by the presence of a thermally insulating layer. Unfortunately, there seems to exist no model for the evolution of surface cratering for asteroids over time, in particular over its early phases, except for the outdated model of \citet{Hou79}, from which one could derive estimates for the evolution of the density of micro-cracks. The model of \citet{Hou79} predicts $>50$ Ma for evolution of a significant surface cratering which seems to be grossly in line with the model results in \citet{Dav13}. Therefore it seems questionable whether the micro-structure of present-day meteorites is representative for the micro-structure of the surface material of planetesimals during the crucial thermal evolutionary phase of the first $\sim100\dots200$ Ma, where temperatures drop below the closure temperatures of the thermochronological systems. 

In view of this uncertainty we consider the effect that a surface layer with a reduced heat conductivity by micro-cracks has on the thermal evolution model by using in Eq.~(\ref{HeatCondPor}) for $K_\mathrm{b}(T)$ the heat conductivity for impact damaged material for some radius range between the surface and some depth $d$ below the surface. For calculating this conductivity we assumed a reduction factor of $s=30$ and a value of the aspect ratio of the micro-cracks of $\alpha=25$, see \citet{Gai18}. The heat conductivity for this case is shown by the blue line in Fig.~\ref{FigKheat}a. The value of $d$ is determined such that we obtain an as good a fit as possible of model results as compared to the meteoritical record.

\subsection{Sintering}
\label{SectApproxSinter}

The porous granular mixture of minerals and some metal from which the parent body of the meteorites of an ordinary chondrite group initially is composed of is compacted to a state with vanishing void space by creep processes and surface diffusion at elevated temperatures and/or high pressures (sintering). At the same time its mineral composition is equilibrated. We include the process of compaction in our thermal evolution model of the L chondrite parent body in the same way as described in \citet{Gai15} for the case of the evolution of the H chondrite parent body. The calculation is based on a theory of isostatic hot pressing used for the modelling of sintering in technical processes. All necessary equations and input data are described in \citet{Gai15}. We have only to observe that the average chondrule diameter in L chondrites is with 0.5 mm bigger than the average diameter of 0.3 mm for chondrules in H chondrites. Because the time required for sintering a granular material strongly increases with the size of its granular units and with decreasing temperature this means that the compaction of chondrules in L chondrite material requires higher temperature than for chondrules in H chondrite material to compensate for the bigger size. 

\begin{figure}

\includegraphics[width=\hsize]{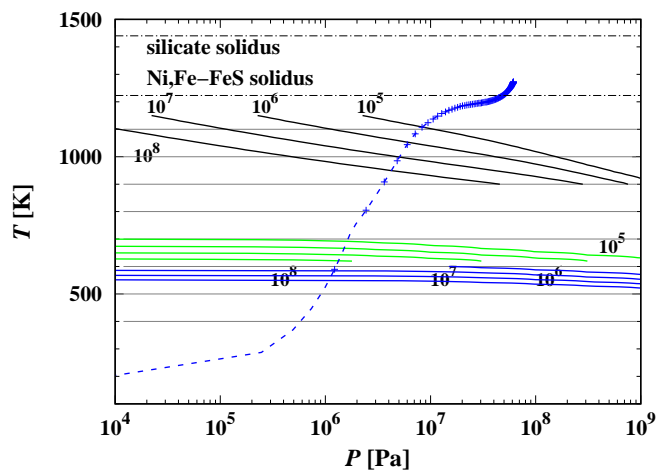}

\caption{Lines of constant time required for compaction of chondritic material of H chondrites at different temperature and pressure conditions. The upper group of lines correspond to the sintering of chondrules with 0.5 mm diameter. The middle group corresponds to sintering of matrix material with particle diameters of 2\,$\mu$m. The numbers at the lines show the time required for compaction (in Ma). The lower group corresponds to sintering of fine dust material with particle diameters of 0.2\,$\mu$m. The blue dashed line corresponds to the highest temperatures achieved at some radius during the evolution of model LM3 for the parent body of the L chondrites and the pressure at that depth. The crosses correspond to radial positions with 1.5 km distance from each other. The horizontal dash-dotted lines correspond to the incipient melting of a Fe,Ni-FeS eutectic at 1\,223~K for L chondrites \citep{Tom09} and that of silicates at 1\,440~K.
}

\label{FigSinterTime}
\end{figure}

The sintering operates in the temperature range above $600\, \dots\,\gtrsim 1\,000$~K, depending on the kind of material and particle sizes. Since temperatures close to the surface of a body stay always below this threshold temperature there remains an outer layer of considerable thickness with barely compacted porous material. It is important to include the compaction process into the modelling of the thermal evolution of the body because from the compaction at high temperatures there results a structure of the body where an extended compacted core with high heat conductivity is mantled by a layer of residual in-compacted material with low heat conductivity. Since the study of \citet{Yom84} on compaction of parent bodies of chondrites and the more since the paper of \citet{Akr98} it is clear that this core-mantle structure strongly modifies the thermal structure of such bodies. In such a case one has a nearly isothermal core and a surface layer with a rather strong temperature gradient, as opposed to the case where heat conductivity is constant or varies only gradually across a body in which case the temperature varies uniformly between surface and centre. 

The variation of porosity $\Phi$ of a granular material is described by Eq.~(\ref{EqSint}) where the right hand side contains contributions from of different mechanisms resulting in a compactification of the material \citep[see][ for details]{Gai15}. For simplicity we consider in the model calculation only the contribution of creep process which is responsible for the deformation and compaction of the chondrule aggregate. 

We also consider in part of the calculations an approximation where we do not model the sintering process explicitly, as it is done for part of our model calculation, but use the simple prescription that 
\begin{equation}
\Phi=\begin{cases}
\Phi_0 &\mathrm{if}\ T_\mathrm{max}\le T_\mathrm{sl}
\\[.5cm]
\displaystyle {T_\mathrm{sh}-T\over T_\mathrm{sh}-T_\mathrm{sl}}\,\Phi_0 & \mathrm{if}\ T_\mathrm{sl}<T_\mathrm{max}\le T_\mathrm{sh}
\\[.5cm]
0&\mathrm{if}\ T_\mathrm{sh}\le T_\mathrm{max}
\end{cases}
\label{EqApprSintPor}
\end{equation} 
where $T_\mathrm{max}$ is the maximum temperature experienced by a mass-element during its past history. This saves much computing time. The prescription is guided by experience with complete numerical simulations of the sintering process \citep{Gai15} that in models for chondrite parent bodies the loss of porosity by sintering occurs over a rather narrow temperature interval. The prescription mimics this by switching in a continuous fashion between the limit cases of low-conductivity for porous material and high-conductivity for compacted material within the temperature interval from $T_\mathrm{sl}$ to $T_\mathrm{sh}$. A similar approach has already be used by \citet{Hev06}.

To justify this approximation, Fig.~\ref{FigSinterTime} shows lines of constant time required for sintering of chondritic material from an initial state of random close packing with $\Phi\approx0.34$ to a state with $\Phi=0.05$ where the pore space looses its connectivity. The lines are constructed by integrating the equations for porosity evolution during hot pressing given in \citet{Gai15} from the initial to the final state at constant $T$ and $P$ for chondrules, matrix material, and for fine dust material. Most important for us is the compaction of the chondrules which occurs by creep processes. The figure also shows the pressure-temperature stratification for the optimised model of the parent body of the L chondrites. This line intersects the line for a duration of the sintering process of 10$^6$ years at a temperature of $\sim1\,060$~K. This duration corresponds to the width of the temperature peak during the time evolution of temperature in the surface-near layers which is essentially ruled by the decay-time of $^{26\!}$Al.

The matrix re-crystallizes and sinters already at lower pressures and temperatures by surface diffusion \citep{Gai15}. According to our sinter model the initially fine grained matrix starts to coarsen and sinter at temperatures between 520 and 550 K, see Fig.~\ref{FigSinterTime}. This does not result in a significant compaction of the material, but in baking together the chondrule-matrix mixture into a sandstone-like material which shows significant cohesion, different to the sand-like character of the material if no sintering of the matrix material occurred. The effective porosity of the mixture would change if matrix material in the contact region between chondrules sinters, but the effect must be small for the parent bodies of ordinary chondrites because of the low matrix abundance ($\sim10$ vol\%). We neglect this. 

In our model calculation we chose the following constants in Eq.~(\ref{EqApprSintPor}): $T_\mathrm{sl}=1\,030$\,K and $T_\mathrm{sh}=1\,090$\,K for chondrule sintering and $T_\mathrm{sl}=540$\,K and $T_\mathrm{sh}=560$\,K for matrix sintering.

\end{appendix}


\end{document}